\documentclass{aa}  

\usepackage{soul}
\usepackage{xcolor}
\usepackage{graphicx}
\usepackage{multirow}
\usepackage{hhline}
\usepackage{amsmath}
\usepackage{xcolor}
\usepackage{adjustbox}

\usepackage{txfonts}

\usepackage{hyperref}
\hypersetup{
    linkcolor=blue,
    citecolor=blue,
    colorlinks=true,
    filecolor=blue,      
    urlcolor=blue
    }

\usepackage{float}
\usepackage{cancel}

\usepackage{natbib}
\bibpunct{(}{)}{;}{a}{}{,}

\begin{document} 

   \title{Velocity fields and turbulence from cosmic filaments to galaxy clusters}

   \titlerunning{Turbulence in Virgo's filaments}
   
   \author{Th\'eo Lebeau\inst{1} \thanks{E-mail: theo.lebeau@universite-paris-saclay.fr}, Saleem Zaroubi\inst{2,3}, Nabila Aghanim\inst{1}, Jenny G. Sorce\inst{4,1}, Mathieu Langer\inst{1}}
   \authorrunning{Lebeau et al.}

   \institute{Université Paris-Saclay, CNRS, Institut d'Astrophysique Spatiale, 91405 Orsay, France
        \and 
            Astrophysics Research Center of the Open University (ARCO), The Open University of Israel, Israel
        \and 
            Kapteyn Astronomical Institute, University of Groningen, Groningen, The Netherlands
        \and
             Université de Lille, CNRS, Centrale Lille, UMR 9189 CRIStAL, F-59000 Lille, France}

   \date{Received 16 January 2025 / Accepted 06 October 2025}
 
  \abstract
  {Galaxy clusters are currently the endpoint of the hierarchical structure formation; they form via the accretion of dark matter and cosmic gas from their local environment. In particular, filaments contribute grandly by accreting gas from cosmic matter sheets and underdense regions and feeding it to the galaxy clusters. Along the way, the gas in filaments is shocked and heated, which, together with the velocity structure within the filament, induces swirling and, thus, turbulence. In this work, we study a constrained hydrodynamical simulation replica of the Virgo cluster at redshift $z=0$ to characterise the velocity field in the two cosmic filaments connected to the cluster with unprecedented high resolution. First, we conduct a qualitative examination of slices extracted from the simulation. We study the temperature, the velocity field, and derived quantities in longitudinal cuts to study the general structure of the filaments and in transverse cuts to study their inner organisation and connection to cosmic matter sheets and underdense regions. Then, we conduct a quantitative study of velocities in Virgo's filaments by computing the 2D power spectrum from 1 and 5~Mpc square maps extracted from the slices and centred on the core of the filaments. We show that the velocity field goes from mostly compressive far in the filaments to mostly solenoidal in Virgo's core. Moreover, we observe that the total power spectrum in the filaments gains in amplitude and steepens towards Virgo.} 


   \maketitle

\section{Introduction}

The hierarchical formation of large-scale structures in the Universe is driven by gravity, in competition with cosmic expansion (e.g. \citeauthor{peebles2020large} \citeyear{peebles2020large}), leading to the anisotropic collapse of primordial density field fluctuations \citep{zel1970gravitational}. Dark matter (DM) and baryons thus flow from underdense, i.e. voids, to overdense environments, i.e. galaxy clusters, connected together by cosmic filaments, altogether forming the cosmic web \citep{bond1996filaments}. At the nodes of this network, galaxy clusters are the most massive gravitationally bound structures in the Universe; they are thus used to constrain, in particular, the cosmic matter density and distribution parameters, i.e. $\Omega_m$ and $\sigma_8$, of the $\Lambda$CDM model. However, to constrain these parameters via the cluster number counts \cite[e.g.][]{planck2014szclustercount,salvati2018constraints,2024A&A...690A.238Aymerich} or the baryon fraction \citep[e.g.][]{wicker2023constraining}, we need to estimate and calibrate their mass precisely and in an unbiased manner (see e.g.\citeauthor{nagai2007effects} \citeyear{nagai2007effects}, \citeauthor{2009ApJ...705.1129Lau} \citeyear{2009ApJ...705.1129Lau}, \citeauthor{2013ApJ...777..151Lau} \citeyear{2013ApJ...777..151Lau}, \citeauthor{biffi2016nature} \citeyear{biffi2016nature}, \citeauthor{salvati2019mass} \citeyear{salvati2019mass}, \citeauthor{gianfagna2021exploring} \citeyear{gianfagna2021exploring}, \citeauthor{lebeau2024amass} \citeyear{lebeau2024amass}, \citeauthor{aymerich2024cosmological} \citeyear{aymerich2024cosmological}, \citeauthor{2025MNRAS.536.3784Braspenning} \citeyear{2025MNRAS.536.3784Braspenning}). In particular, when deriving the mass of a cluster from observations of its intracluster medium (ICM) in the X-ray or sub-millimetre wavelengths, we usually assume hydrostatic equilibrium (see \citeauthor{ettori2013mass} \citeyear{ettori2013mass} or \citeauthor{kravtsov2012formation} \citeyear{kravtsov2012formation} for reviews). However, it has been shown in several studies, both in observations and simulations (see, e.g. \citeauthor{salvati2018constraints} \citeyear{salvati2018constraints}, \citeauthor{gianfagna2021exploring} \citeyear{gianfagna2021exploring}, \citeauthor{lebeau2024amass} \citeyear{lebeau2024amass} and references therein) that this hypothesis tends to be too strong, and that in addition with data analysis choices, e.g. accounting for substructures \citep[e.g.][]{meneghetti2010weighing}, or instrumental effects \citep[e.g.][]{2015A&A...575A..30Schellenberger}, there is a biased cluster mass estimation, the so-called hydrostatic mass bias. On the side of physical sources of bias, it was shown that turbulence in the ICM could add from a few per cent to 30\% of non-thermal pressure support \citep[e.g.,][]{rasia2004dynamical,nelson2014hydrodynamic,2018MNRAS.481L.120Vazza,pearce2020hydrostatic,angelinelli2020turbulent}, accounting for it in the mass estimation could contribute in alleviating the hydrostatic mass bias.\\

In the context of precision cosmology, it has therefore become crucial to quantify turbulence in the ICM and understand how it develops. Thanks to the increase in size and resolution of cosmological simulations in the past decade, an important effort has been dedicated to constraining turbulence in the ICM from simulations \citep[e.g.][]{norman1999cluster,2005MNRAS.364..753Dolag,nagai2007effects,gaspari2013constraining,2014ApJ...782...21Miniati,porter2015vorticity,vazza2017turbulence,valles2021troubled,2024A&A...690A..20Ayromlou,2025A&A...693A.263Groth}. It has been shown that vorticity, and then turbulence, is mainly generated by shocks due to mergers \citep[e.g.][]{zuhone2011parameter,vazza2012turbulence,nagai2013predicting}, by accretion of cosmic gas \citep[e.g.][]{iapichino2011turbulence}, and by the feedback of Active Galactic Nuclei (AGN) in cluster cores \citep[e.g.][]{vazza2013thermal,bourne2017agn}, involving multiple injection scales. In addition, the ICM is multiphase \citep[e.g.][]{wang2021non,mohapatra2022characterizing} and stratified \citep[e.g.][]{shi2018multiscale,mohapatra2020turbulence,simonte2022exploring,wang2023turbulent}, and other mechanisms such as galaxy motion \citep[e.g.][]{ruszkowski2011galaxy} and cosmic rays pressure \citep[e.g.][]{2009ApJ...707.1541Beresnyak} could contribute to turbulence. The combination of these processes leads to a more complex turbulence than Kolmogorov's idealised case \citep{kolmogorov1941local}, making its modelling challenging. \\

At the same time, studies of the filament populations based on simulations \citep[e.g.][]{cautun2014evolution,gheller2019,pereyra2020detection,2020MNRAS.494.5473Kuchner,galarraga2021properties,rost2021the300,gouin2022gas}, including their detectability through HI 21cm emission \citep{2017MNRAS.468..857Kooistra,2019MNRAS.490.1415Kooistra}, and observations \citep[e.g.][]{tanimura2022xray,gouin2023soft,2024A&A...692A.200Gallo} have shown the impact of cosmic filaments on cluster shapes and dynamical states \citep[e.g.][]{gouin2020probing,gouin2021shape,2024A&A...692A..44Santoni,2025A&A...698A.201Capalbo} and how cosmic gas is fed to clusters through filaments \citep{2020MNRAS.499.2303Valles,vurm2023cosmic,2022MNRAS.510..581Kuchner}. However, to date, although some works studied shocks and heating of gas in filaments \citep[e.g.][]{ryu2003waves,pfrommer2006shock,ryu2008turb,benett2020shocks}, amplification of magnetic fields in filaments \citep[][]{2014MNRAS.445.3706Vazza}, which is tightly related to turbulence, and also velocity fields \citep[e.g.][]{2010ApJ...712....1Zhu,2013ApJ...777...48Zhu} and generation of turbulence at the interface of the Intergalactic Medium and galaxy halos \citep[e.g.][]{2018A&A...610A..75Cornault}, to our knowledge, none studied deeply how vorticities develop along the way from the filaments connected to cluster to the ICM. The aim of this work is thus to conduct both a qualitative and quantitative study of gas dynamics in filaments attached to a specific cluster, Virgo. \\

To this end, we examine the structure of the velocity field and temperature along large filaments connected to a state-of-the-art high-resolution simulated replica of the Virgo cluster in its local environment \citep{sorce2021hydrodynamical} at redshift $z=0$. In the following, we introduce the simulation and the methodology adopted to construct slices of temperature, velocity, and derived quantities in Sect. \ref{sec:2}. We qualitatively describe the gas flow and the sources of turbulence using longitudinal and transverse cuts of the filaments connected to the Virgo cluster, respectively, in Sect. \ref{sec:3} and \ref{sec:4}. In Sect. \ref{sec:5}, we decompose the velocity field into compressive and solenoidal components and compute their 2D power spectra, as well as that of the total velocity field, in the transverse slices. We discuss our results in Sect. \ref{sec:6} and conclude in Sect. \ref{sec:7}.

\begin{figure*}
        \begin{minipage}[s]{1\textwidth}
            \centering
            \includegraphics[trim=20 250 20 300, width=.99\textwidth]{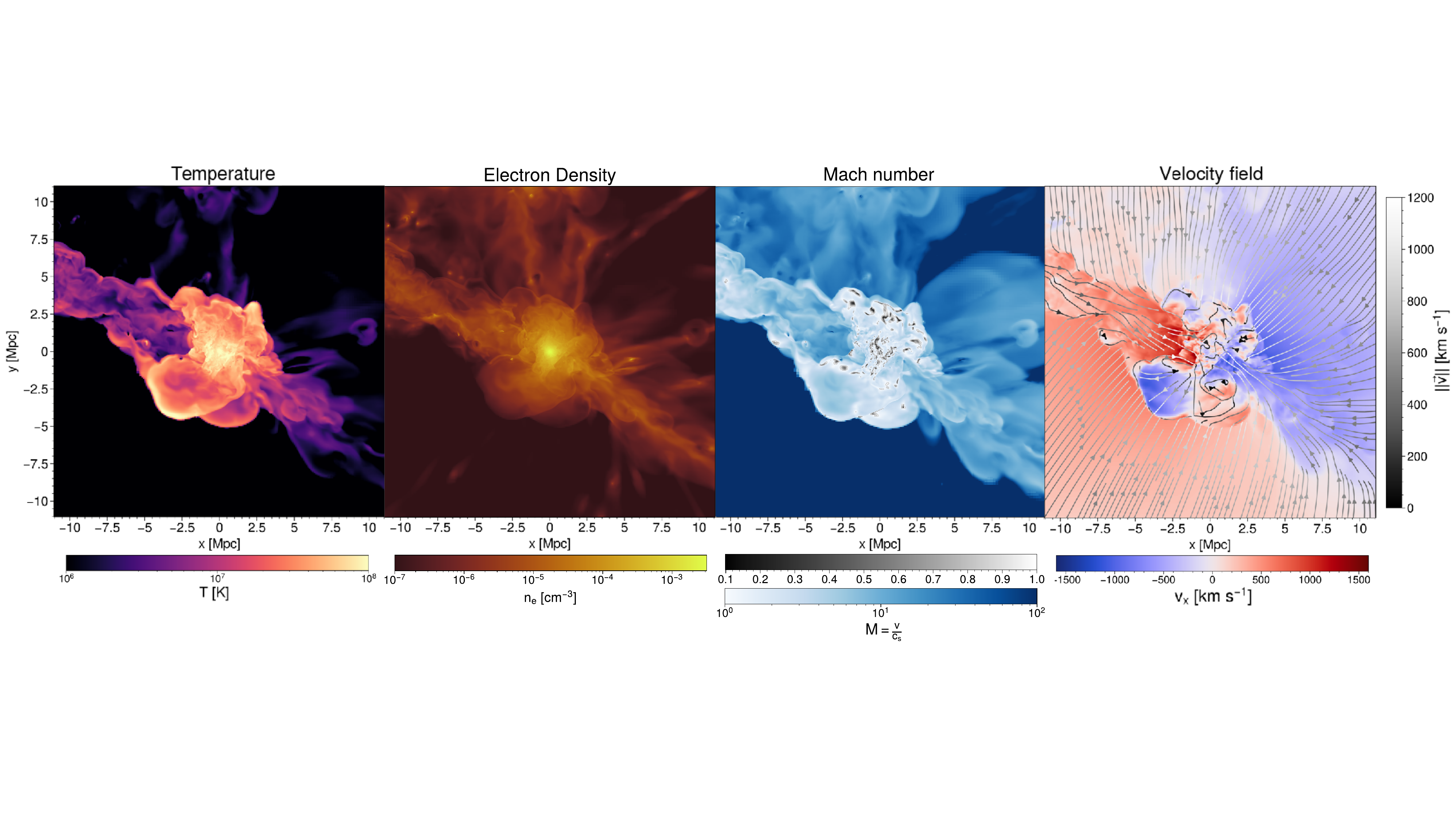}
            \caption{From left to right, longitudinal cuts of temperature, electron density, Mach number and velocity in Virgo replica and the two filaments to which it is connected. The maps are about 22~Mpc wide, contain $\mathrm{15728^2}$ pixels, and are centred on Virgo's core. On the third panel, Mach number, colour bands are used to display the large range of values: the black-to-white is on a linear scale in the [0.1,1] range, and the white-to-blue is on a log scale in the [1,100] range. On the right panel, the background map is the velocity along the x-axis of the simulation box, $v_x$, and the foreground arrows represent the norm of the velocity field in the x-y plane, $||\vec{v}||$. The background (foreground) filament is on the left (right) part of each panel. The buoyant bubble is located roughly at a position $(x=-5.0; y=0.5)$~Mpc. Multiple galaxies are visible on the density map, for example, at positions $(-2.5; -0.5)$ and $(-6.0; 1.0)$~Mpc.}
            \label{long_cut}
       \end{minipage}
\end{figure*}

\section{Methodology}
\label{sec:2}

\subsection{The Virgo replica simulation}
\label{sec:2.1}

Using constrained initial conditions of the local Universe, reconstructed from the position and peculiar velocities of galaxies in the Cosmicflows-2 \citep{tully2013cosmicflows} dataset (see \citeauthor{sorce2016cosmicflows} \citeyear{sorce2016cosmicflows} for details), \cite{sorce2019virgo} simulated DM replicas of the Virgo cluster within its local cosmic environment. Among the 200 DM simulations, the most representative one, in terms of average properties and merging history compared to the full sample, has been used to run a high-resolution zoom-in hydrodynamical simulation of Virgo \citep{sorce2021hydrodynamical}. Both the DM and high-resolution hydrodynamical simulations agree well with observations. In particular, they accurately reproduce the observed filament in the foreground of Virgo along our observation sightline, as well as the group of galaxies falling onto Virgo. More details about the simulation and its agreement with observations can be found in \cite{sorce2021hydrodynamical} and \cite{lebeau2024amass}. \\

This simulation uses the \cite{planckcosmoparam2014} cosmological parameters; the Hubble constant, $H_0=67.77~\mathrm{km~s^{-1}~Mpc^{-1}}$, dark energy density, $\Omega_{\Lambda}=0.693$, total matter density, $\Omega_\text{m}=0.307$,  baryonic density, $\Omega_\text{b}=0.048$, amplitude of the matter power spectrum at 8~$\mathrm{Mpc~h^{-1}}$, $\sigma_8 = 0.829$,  and spectral index, $n_\text{s}=0.961$. It has been produced using the adaptive mesh refinement (AMR) hydrodynamical code {\tt RAMSES} \citep{teyssier2002cosmological}. In this run, Euler equations are solved following the MUSCL-Hancock method using the Riemann solver of \citet{1994ShWav...4...25Toro}, which is second-order accurate in space, with a variable time step based on the Courant Friedrich Levy condition \citep[see][for more details]{teyssier2002cosmological}. Within the 500~$\mathrm{Mpc~h^{-1}}$ local Universe box, the zoom region is a 30~Mpc diameter sphere with an effective resolution, i.e. only in the zoom region, of $8192^3$ DM particles of mass $m_{\mathrm{DM}}=3\times10^7~\mathrm{M_\odot}$. The finest cell size of the AMR grid is 0.35~kpc following a pseudo-Lagrangian refinement criterion, i.e. based on DM particles density \citep[see][for more details]{sorce2021hydrodynamical}. In the high-resolution run, the maximum resolution of 0.35~kpc is reached only in the core of the cluster; in the filaments, the resolution of the cells is mostly 22.5 and 45~kpc. In the following, we used a best resolution of 22.5~kpc to compute the Hodge-Helmholtz decomposition and the power spectra. In addition, it is worth mentioning that the development of turbulent motion at small scales in this simulation is limited by numerical diffusion due to the accuracy order of the Riemann solver and the maximum resolution of the AMR grid. Therefore, the results presented in this study are to be taken as a lower limit of the actual intensity of turbulent motion in the ICM that could be better resolved with a higher-order Riemann solver and a finer grid resolution.\\

In addition, the simulation implements sub-grid models for radiative gas cooling and heating, star formation and kinetic feedback from the AGN and type II supernova (SN) similar to the Horizon-AGN implementation of \cite{dubois2014dancing,dubois2016horizon} with an AGN feedback model improved by orienting the jet according to the black hole (BH) spin (see \citeauthor{dubois2021introducing} \citeyear{dubois2021introducing} for details). The simulation was run from redshift $z=120$ to $z=0$. In this work, we only used the snapshot at $z=0$.\\

\subsection{Longitudinal and transverse cuts along the Virgo filaments}
\label{sec:2.2}

We used the {\tt rdramses} code\footnote{\url{https://github.com/florentrenaud/rdramses}, first used in \cite{renaud2013sub}.}, to extract the DM particles and gas cell properties from the simulation. To identify the Virgo DM halo and its galaxies, we used the \cite{tweed2009building} halo finder, respectively, on the DM particles and the star particles. In this work, the gas is defined as the baryonic component in the simulation cells, in which hydrodynamics equations are solved following an Eulerian approach; it thus does not include the stars. We extracted very thin slices from the simulation to study variations of temperature and velocity at small scales in the filaments connected to Virgo. We adapted the code used in \cite{lebeau2024amass} to do so. We set the integration depth along a given sightline to 22.5~kpc, which is 64 times the finest cell size of 0.35~kpc. In underdense regions, cells can be larger than 22.5~kpc but still partly belong to the selected slice; in that case, only their fraction inside the slice has been accounted for. \\

\begin{figure*}
        \begin{minipage}[s]{1\textwidth}
            \centering
            \includegraphics[trim=180 210 100 150,clip, width=.9\textwidth]{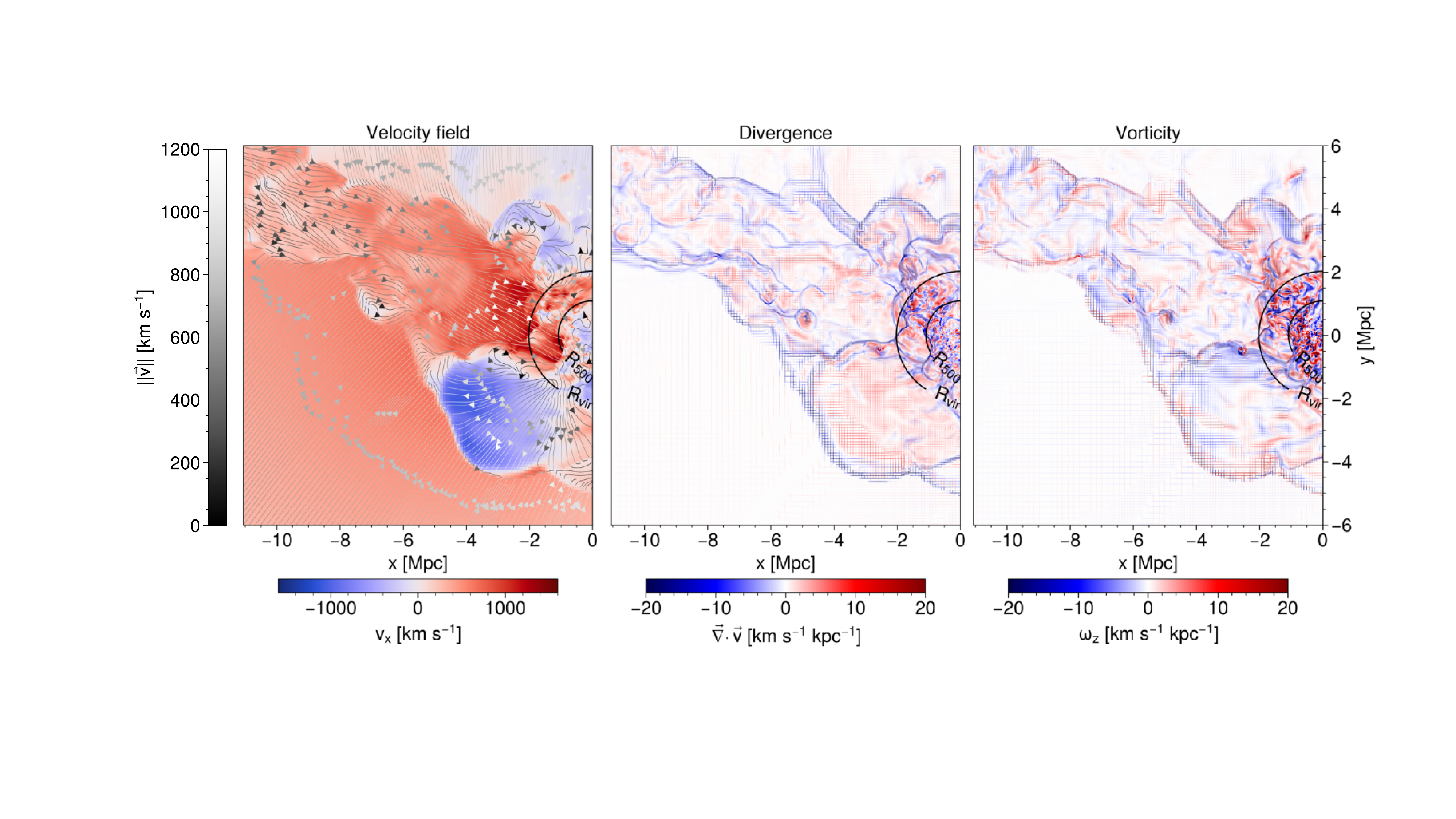}
            \caption{Zoom on the background filament in the longitudinal cut. The maps are $11.061\times12$~Mpc large with the same resolution as those presented in Fig. \ref{long_cut}. The left panel shows the velocity field, similar to the right panel of Fig. \ref{long_cut}, the central panel shows the divergence, and the right panel shows the z component of the vorticity. Virgo's $R_{500}$ and virial radius, $R_{\mathrm{vir}}$, are shown as black circle arcs on each map. The merger-accelerated accretion shock is located roughly in the range x=[-5,-2]~Mpc and y=[-5,-1]~Mpc.}
            \label{zoom_left}
       \end{minipage}
\end{figure*}

\begin{figure*}
        \begin{minipage}[s]{1\textwidth}
            \centering
            \includegraphics[trim=100 120 20 200,clip, width=.9\textwidth]{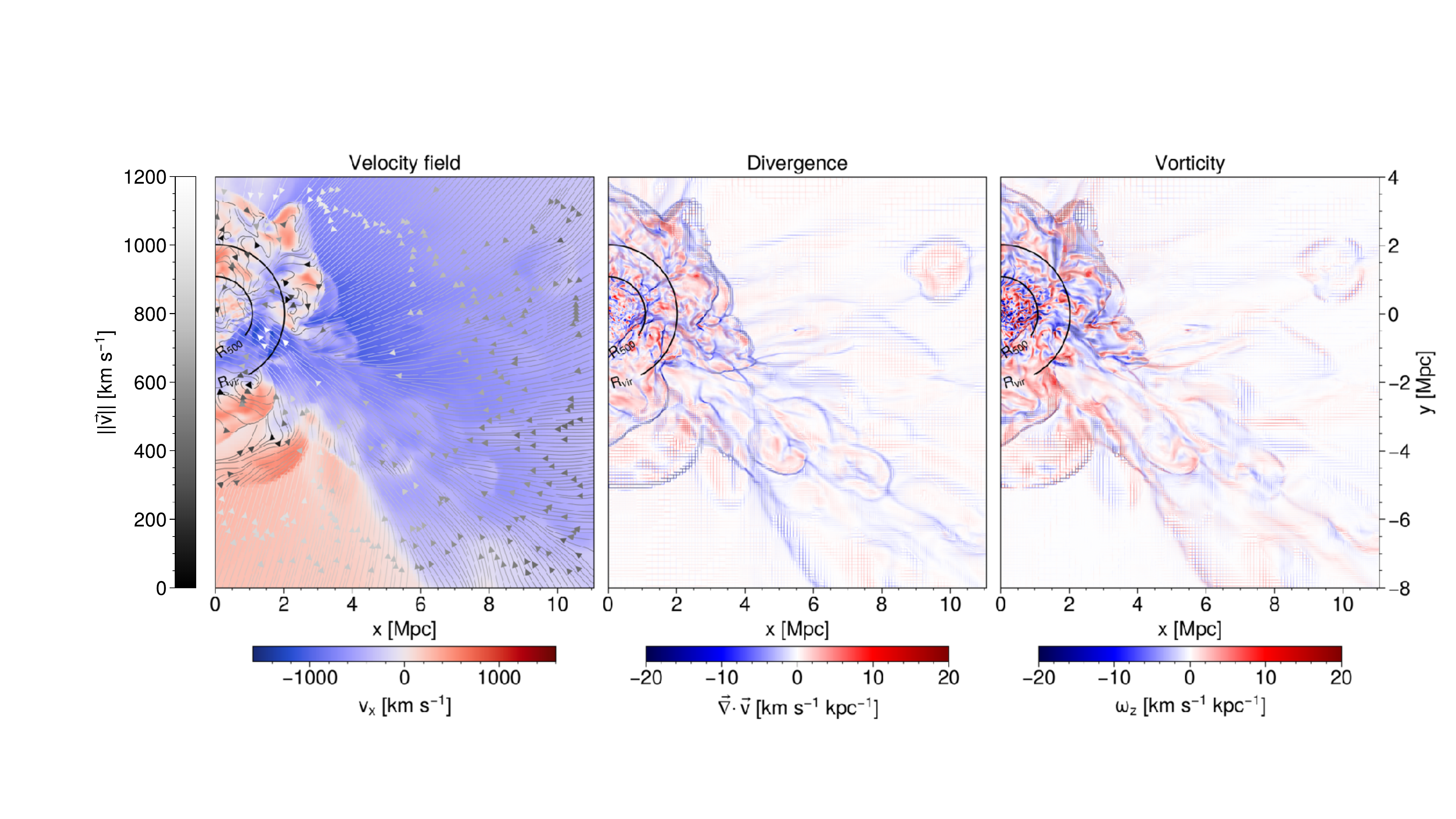}
            \caption{Zoom on the foreground filament in the longitudinal cut. The maps are $11.061\times12$~Mpc large with the same resolution as those presented in Fig. \ref{long_cut}. The left panel shows the velocity field, similar to the right panel of Fig. \ref{long_cut}, the central panel shows the divergence, and the right panel shows the z component of the vorticity. Virgo's $R_{500}$ and virial radius, $R_{\mathrm{vir}}$, are shown as black circle arcs on each map.}
            \label{zoom_right}
       \end{minipage}
\end{figure*}

The Virgo replica is connected to two cosmic filaments aligned along the same direction, almost perfectly in the $xy$ simulation box plane. We thus extracted a slice in this plane centred on the Virgo cluster core; we call it the "longitudinal cut" throughout the paper; it is presented in Fig. \ref{long_cut}. Moreover, we show a zoom on the filament with $\Delta x_{\mathrm{cen}}=x-x_{\mathrm{virgo\,centre}}<0$ ($\Delta x_{\mathrm{cen}} >0$) in Fig. \ref{zoom_left} (Fig. \ref{zoom_right}), $x_{\mathrm{virgo\,centre}}$ being the centre of the Virgo DM halo. Given that this simulation replicates the Virgo cluster in its local environment, including the Milky Way, we can put ourselves in the sightline of the observer, as in \citet{lebeau2024amass}. Thus, we can call them, respectively, "background filament" (left side of panels of Fig. \ref{long_cut}) and "foreground filament" (right side of panels of Fig. \ref{long_cut}).  \\

\begin{figure*}[h!]
        \begin{minipage}[s]{1\textwidth}
            \includegraphics[trim=150 150 150 0, width=1\textwidth]{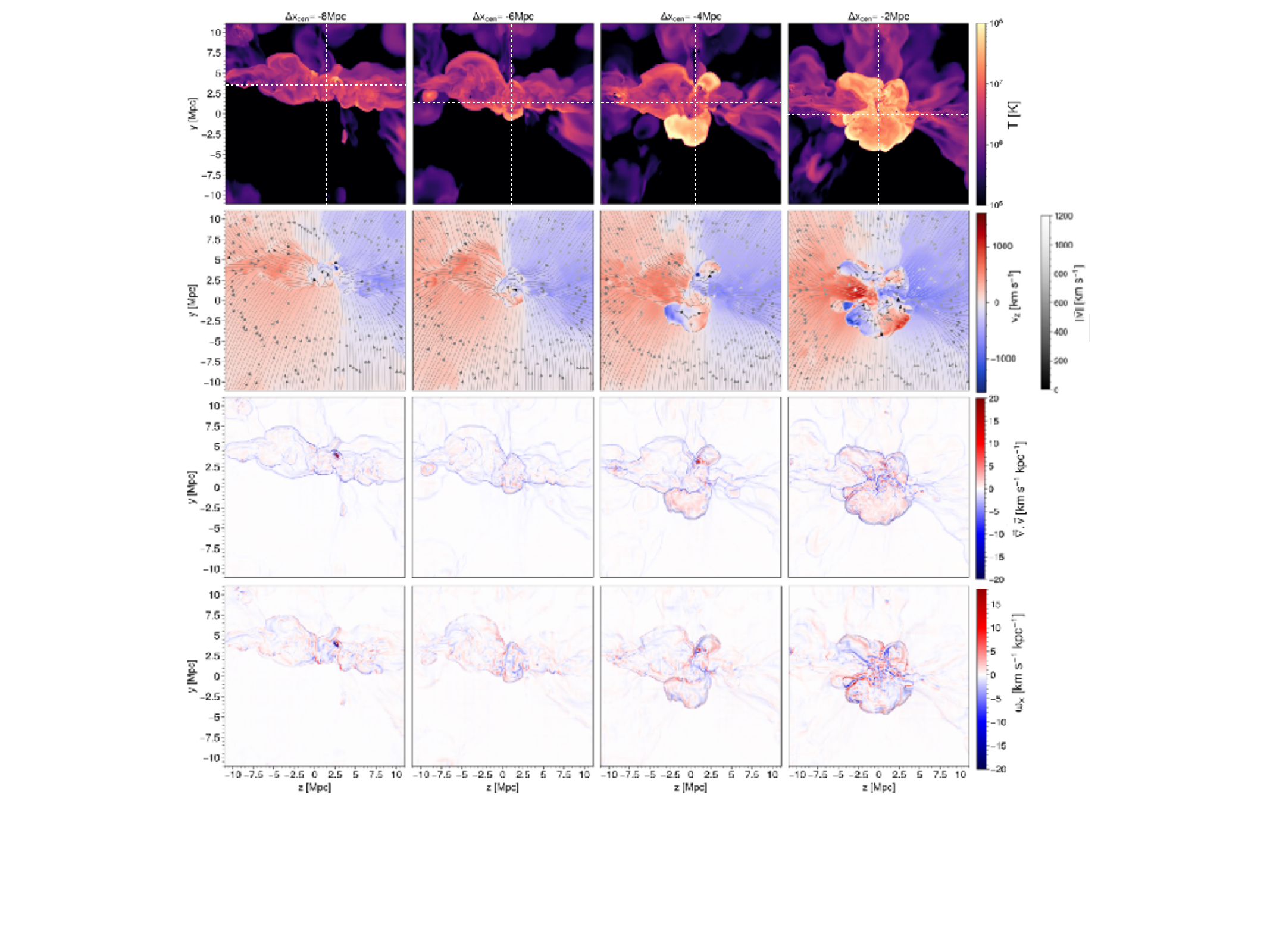}
            \caption{Transverse cuts along the background filament. From left to right, we present cuts at distances of 8, 6, 4 and 2~Mpc from Virgo's centre. From top to bottom, we present the temperature, the velocity field with the background map being the velocity along the $z$ simulation box axis, $v_z$, and the foreground arrows representing the norm of the velocity field in the $zy$ plane, $||\vec{v}||$, the divergence of the velocity field and the $x$ component of its vorticity. Similarly to longitudinal cuts, the maps are 22.122 Mpc wide, contain $\mathrm{15728^2}$ pixels, and are centred on Virgo's centre.}
            \label{left_fil_t_v_div_rot}
       \end{minipage}
\end{figure*}

\begin{figure*}[h!]
        \centering
        \begin{minipage}[s]{1\textwidth}
            \centering
            \includegraphics[trim=200 10 140 10, clip, width=0.85\textwidth]{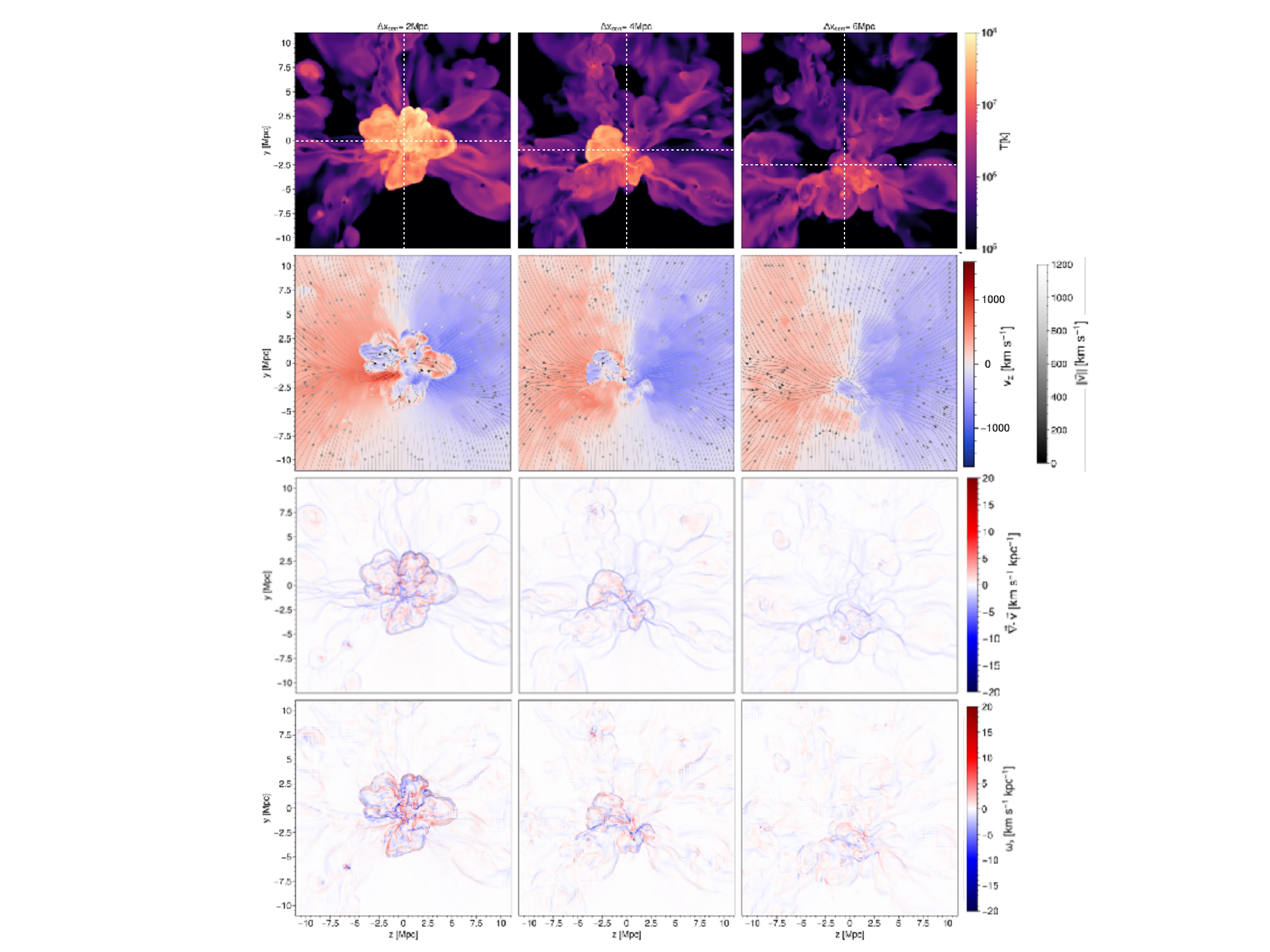}
            \caption{Transverse cuts along the foreground filament. From left to right, we present cuts at distances of 2, 4 and 6~Mpc from Virgo's centre. From top to bottom, we present the temperature, the velocity field with the background map being the velocity along the $z$ simulation box axis, $v_z$, and the foreground arrows representing the norm of the velocity field in the $zy$ plane, $||\vec{v}||$, the divergence of the velocity field and the $x$ component of its vorticity. Similarly to longitudinal cuts, the maps are 22.122 Mpc wide, contain $\mathrm{15728^2}$ pixels, and are centred on Virgo's centre.}
            \label{right_fil_t_v_div_rot}
       \end{minipage}
\end{figure*}

Then, we extracted perpendicular slices to the filaments to study the temperature and velocity fields in their core. We extracted slices in planes parallel to the $xz$ simulation box plane instead of computing slices with a normal vector aligned with the major axis of the filaments (the angle between the two is roughly 25 degrees) to avoid non-physical small-scale blurring due to the misalignment with the AMR grid. This choice preserves the resolution at very small scales, allowing a qualitative analysis of the structure of both temperature and the velocity fields in the filaments. These slices are presented in Figs. \ref{left_fil_t_v_div_rot} and \ref{right_fil_t_v_div_rot} for the background and foreground filaments, respectively. To visualise the position of the slices, we refer the reader to Fig. \ref{app:3D_vis} in Appendix \ref{appendix 3D vis}. \\

The slices presented in Fig. \ref{long_cut}, \ref{left_fil_t_v_div_rot} and \ref{right_fil_t_v_div_rot} are $(22.122~\text{Mpc})^2$ and contain $15728^2$ pixels; each pixel is thus 1.4~kpc wide. The pixel size is the same for the zooms presented in Figs. \ref{zoom_left} and \ref{zoom_right}. For each of the longitudinal and transverse cuts, we present the mass-weighted mean temperature and electron density (the transverse cuts can be found in Appendix \ref{appendix ne transverse}), and the mean velocity of the cells integrated along the sightline in the Virgo restframe. For the velocity, the blue-to-red field in the background is the velocity along the abscissa of the figure, that is, $v_x$ ($v_z$) for the longitudinal (transverse) cut in Figs. \ref{long_cut}, \ref{zoom_left} and \ref{zoom_right} (Figs. \ref{left_fil_t_v_div_rot} and \ref{right_fil_t_v_div_rot}), and the black-to-white arrows in the foreground represent the velocity field in the plane of the slice. In addition, we present the 2D divergence and the $z$ ($x$) component of the curl, that is, the vorticity of the velocity field in the longitudinal (transverse) cuts on the central and right panels of Figs. \ref{zoom_left} and \ref{zoom_right} (third and bottom panels of Figs. \ref{left_fil_t_v_div_rot} and \ref{right_fil_t_v_div_rot}). For the longitudinal cut, the divergence in the $xy$ plane and the $z$ component of the vorticity are defined as
\begin{equation}
    \Vec{\nabla}\cdot\Vec{v} = \frac{\partial{v_x}}{\partial{x}}+\frac{\partial{v_y}}{\partial{y}} \quad \textrm{and} \quad \omega_z = \frac{\partial{v_y}}{\partial{x}}-\frac{\partial{v_x}}{\partial{y}}.
\end{equation}

\noindent For the transverse cut, the divergence in the $yz$ plane and the $x$ component of the vorticity are defined as

\begin{equation}
    \Vec{\nabla}\cdot\Vec{v} = \frac{\partial{v_y}}{\partial{y}}+\frac{\partial{v_z}}{\partial{z}} \quad \text{and} \quad \omega_x = \frac{\partial{v_z}}{\partial{y}}-\frac{\partial{v_y}}{\partial{z}}.
\end{equation}

In Sect. \ref{sec:5}, we decompose the velocity field in compressive and solenoidal components following the Hodge-Helmholtz procedure as in \citet{vazza2017turbulence}. However, this can only be done on a regular 3D grid, whereas our simulation of Virgo, based on the AMR code RAMSES, has an irregular grid. We thus adapted the simulation output data specifically for this section. Given the range of the turbulent scales considered in this study, roughly from a few Mpc to a few tens of kpc, we did not need to use the simulation at its finest resolution to study statistical properties of the velocity field through the power spectrum; we thus set the maximum resolution level to cells of 22.5~kpc while reducing the output data with {\tt rdramses}. It means that we did not use smaller cells than this resolution limit in the AMR dependence tree; the outputted physical properties are thus the mean of all the smaller "children" cells, as defined in \cite{teyssier2002cosmological}, associated with a given larger "parent" cell. It is worth noting that this cell size equals the thickness of the slices considered in this work. Cells with a lower resolution, i.e. larger cells in underdense regions, were divided into smaller ones up to the resolution limit with the same physical properties except for the mass equally distributed among all the new smaller cells to keep the same density. Choosing this resolution limit also avoided generating a large amount of data, given that the number of cells is multiplied by eight each time we increase the resolution level by one. It also significantly reduced the calculation time. \\ 

\section{Matter flows across the filaments}
\label{sec:3}

In this section, we focus on the longitudinal cut to study the gas across the filament. \\

First, on the left panel of Fig. \ref{long_cut}, the temperature reveals the components of Virgo's local environment. On the left part of the slice, a filament of almost 5~Mpc diameter extends over more than 10~Mpc from the limit of the simulation box to very deep in Virgo. The temperature in the core of the filament is in the range $\mathrm{[10^{6.5},10^7]~K}$, with an electron density below $10^{-4}$ $\mathrm{cm^{-3}}$, it is thus a Warm-Hot Intergalactic Medium (WHIM) given the classification in the literature \citep[e.g.][]{2019MNRAS.486.3766Martizzi, gouin2022gas}. At the borders of the filament, the temperature can reach up to $\mathrm{10^{7.5}~K}$ due to the filament accretion shocks, as discussed below. On the right part of the slice, we observe a smaller, hotter and less collimated filament connected to Virgo. Then, underdense regions surround Virgo and the filaments, with temperatures below $\mathrm{10^6~K}$. Finally, the temperature in Virgo's ICM ranges from $\mathrm{10^7~K}$ to more than $\mathrm{10^8~K}$, as expected for galaxy clusters. \\

This thin slice of  22.5~kpc width reveals small-scale temperature fluctuations in Virgo's filaments and core. In the background filament, numerous wavy structures are visible; these might be the signature of instabilities leading to rotation and potentially turbulence. They might be generated by Kelvin-Helmholtz instabilities due to streams with different relative velocities and by galaxies acting as obstacles generating trailing turbulence. In Virgo's core, temperature fluctuations are even at smaller scales and show a very complex gas stirring, which traces complex turbulence. \\

On the second panel, we present the electron density. The densest region is Virgo's ICM, as expected due to the gravitational potential well. Then, in the filaments, by zooming in on the density map, we observe elongated density peaks with a size of about a hundred kpc, coinciding with cold regions in the temperature maps. These cold density peaks are, in fact, galaxies, given that we could match their positions with the positions of galaxies detected by the halo finder. For example, two galaxies are positioned roughly at the coordinates $(x=-2.5;y=-0.5)$ and $(-6.0; 1.0)$~Mpc. On the other hand, we observe an underdense and hot region of similar scale in the background roughly at coordinates $(-5.0; 0.5)$~Mpc; this buoyant bubble might be due to SN or AGN feedback. \\

Then, on the central panel in Fig. \ref{long_cut}, we present the dimensionless flow Mach number, $M=v/c_\text{s}$, with $v$ the magnitude of the velocity field and $c_\text{s}$ the sound speed defined as
\begin{equation}
    c_\text{s}=\sqrt{\frac{\gamma k_\text{B} T}{\mu m_\text{p}}},
    \label{sound speed}
\end{equation}

\noindent with $k_\text{B}$ the Boltzmann constant, $\mu$ the mean molecular weight, $m_\text{p}$ the proton mass and $\gamma$ the adiabatic index set to $5/3$ in the simulation, given that we assume for simplicity a monoatomic perfect gas. We use two colour bands to display the large range of values: the black-to-white is on a linear scale in the [0.1,1] range, and the white-to-blue is on a log scale in the [1,100] range. \\

Similarly to previous works \citep[e.g.][]{gaspari2013constraining,vazza2017turbulence}, we can see that the gas flow is subsonic to transonic in the ICM with a flow Mach number below $M=2$. Then, in the filaments, the gas flow is supersonic, between $M=2$ and $M=20$, which is due both to the temperature and density drop and to the higher overall velocity in the filament, as we can see on the right panel of Fig. \ref{long_cut}. In the upper underdense region, where the temperature is lower, but the density is not much lower than in the filaments, we found flow Mach numbers similar to the latter. Finally, in the lower underdense region, where the density and temperature are much lower than in the upper one, the flow Mach number is of the order of $M=100$, which is consistent with \cite{ryu2003waves}. This is because the low temperature in this medium induces a small sound speed, thus increasing the flow Mach number for a velocity similar to that in the upper underdense region; it is thus a highly supersonic flow. \\

Finally, we focus on the velocity field in the slice presented on the right panel of Fig. \ref{long_cut}. We observe that most of the cosmic gas is first accreted from underdense regions to the cosmic matter sheet, particularly around the background filament (see Sect. \ref{sec:3}), then to filaments and finally to Virgo. It is accelerated up to 1000~$\mathrm{km~s^{-1}}$ in the underdense regions and matter sheet before being slowed down and heated when entering the filament and then re-accelerated up to 1200~$\mathrm{km~s^{-1}}$ before being strongly shocked and heated again, at the termination shock as called by, e.g., \citet{vurm2023cosmic} when entering the cluster. We see how filaments funnel the gas to Virgo and act as cosmic highways, as shown in, e.g., \citet{gouin2022gas} and \cite{vurm2023cosmic}. Virgo is thus the accretion endpoint where the two major flows from the background and foreground filaments meet and stir. Indeed, the $x$ component of the velocity field, $v_x$, is entirely positive (red) in the background filament. In contrast, it is entirely negative (blue) in the foreground filament, and there is a mix of pale red and blue in Virgo. Finally, a strong outflow is located below Virgo on this slice, located roughly in the range x=[-5,-2]~Mpc and y=[-5,-1]~Mpc, which seems to generate vorticity at its boundaries; see left panel of Fig. \ref{zoom_left} for more details. This outflow has already been identified in \cite{lebeau2024can} and might result from the recent major merger identified in the simulation by \cite{sorce2021hydrodynamical}. It ressembles a merger-accelerated accretion shock as studied in \citet{2020MNRAS.494.4539Zhang-MA_accretion_shock}, which could generate vorticity via the baroclinic mecanism \citep[see e.g.][]{vazza2017turbulence,2017MNRAS.471.3212Wittor}.\\

To highlight the velocity field structure in the filaments, particularly shocks and rotation, we show zooms of the background and foreground filaments of the velocity field, their divergence and the vorticity, respectively, in Figs. \ref{zoom_left} and \ref{zoom_right}. The Mach number and density-weighted divergence and vorticity in these zoom regions can also be found in Appendix \ref{appendix zoom}. On each map, Virgo's $R_{\mathrm{500}}=1.09$~Mpc and virial radius, $R_{\mathrm{vir}}=2.15$~Mpc, are shown as black circle arcs. For the background filament, in Fig. \ref{zoom_left}, we observe that the divergence (central panel) highlights the filament boundary. Before entering the filament, the gas flow is essentially laminar with a velocity between 800 and 1000 $\mathrm{km~s^{-1}}$ in a single coherent flow, and the divergence is null. When entering the filament, it experiences an oblique shock, or filament accretion shock, as shown by the divergence, and is thus slowed down and heated, as shown in Fig. \ref{long_cut}. Moreover, the gas starts swirling, as shown by the vorticity (right panel). Then, in the filament, the gas flow is accelerated by its collapse on Virgo, from less than 200 $\mathrm{km~s^{-1}}$ at $\Delta x_{\mathrm{cen}}=-10$~Mpc to more than 1200 $\mathrm{km~s^{-1}}$ at $\Delta x_{\mathrm{cen}}=-2$~Mpc. It is a more complex flow with multiple parallel streams, as shown by the divergence that transports a mildly swirling gas with relatively large eddies from $\Delta x_{\mathrm{cen}}=-10$~Mpc to $\Delta x_{\mathrm{cen}}=-2$~Mpc. Finally, the gas enters the cluster; there is an orthogonal shock, named termination shock in, e.g.,\citet{vurm2023cosmic}, at $\Delta x_{\mathrm{cen}}=-2$~Mpc, which is of about the virial radius, as shown by the high divergence at this distance. The gas is once again slowed down and heated. We also observe that in Virgo's core, the gas highly rotates at much smaller scales, partly due to this termination shock. \\

The picture is even more complex for the foreground filament, presented in Fig. \ref{zoom_right}. We observe a gas flow, but the filament from the bottom right corner of the map to Virgo's core is less distinctly defined. It is visible on the divergence map as this tracer shows multistreams inside the filament, but its boundaries are not as clear as for the background filament. It is confirmed by the vorticity, where we observe slow and large-scale rotation in the same region, and by the temperature on the left panel of Fig. \ref{long_cut}. Nonetheless, we observe the same features as for the background filament: the gas is shocked twice, first when entering the filament and then in the cluster close to the virial radius. This generates eddies, first quite slow and at large scales in the filament, then much quicker and at small scales in the cluster. \\

From this longitudinal cut analysis, the picture that emerges is that Virgo is fed by two aligned filaments with different structurations, the collimated background filament embedded in a cosmic sheet and the funnel-shaped foreground one, but comparable gas flows, both contributing to the complex gas stirring in Virgo's core.

\section{Matter flows from cosmic matter sheet to filaments }
\label{sec:4}

\begin{figure*}
        \begin{minipage}[s]{1\textwidth}
            \centering
            \includegraphics[trim = 0 280 0 0, width=1\textwidth]{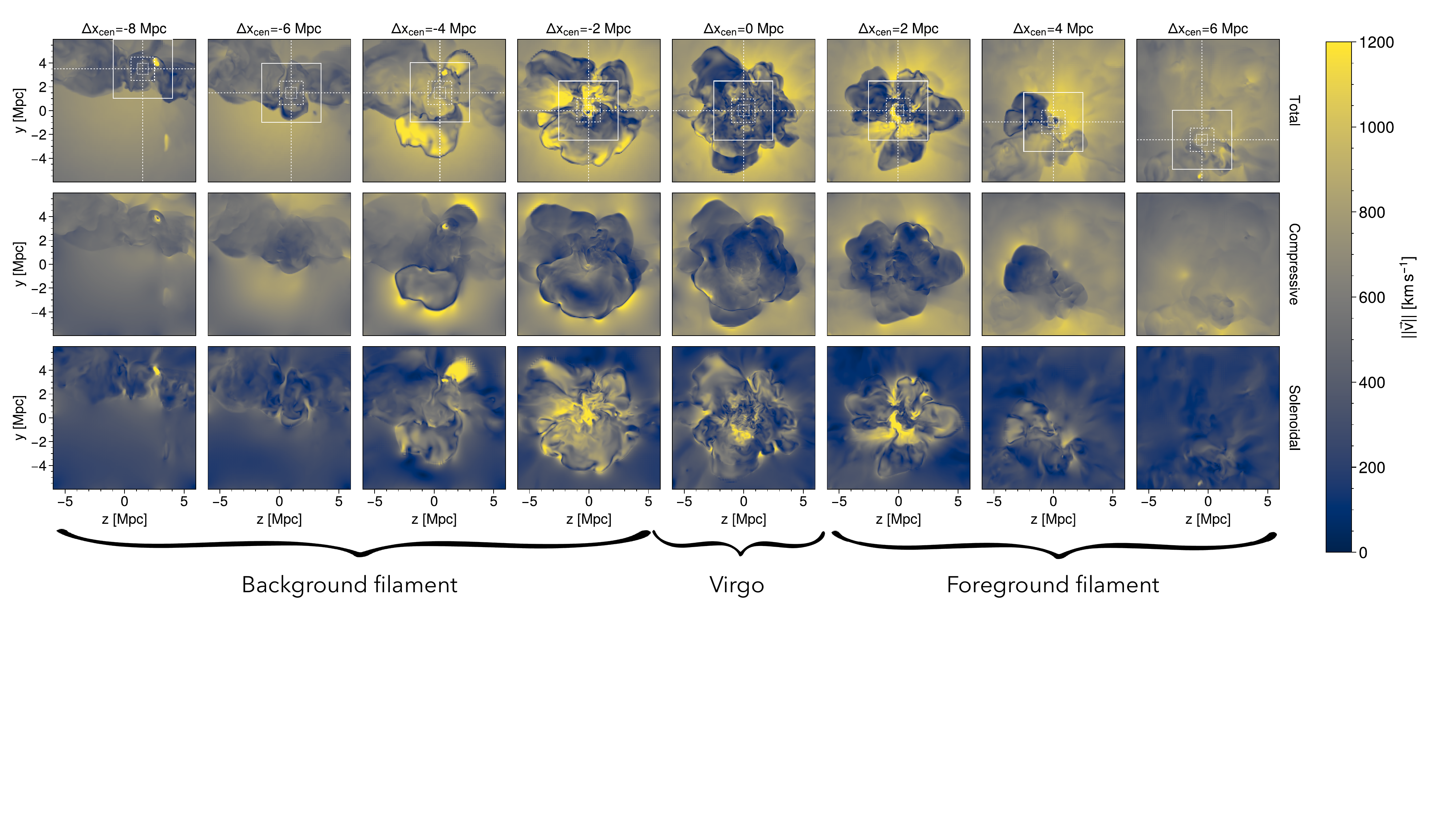}
            \caption{Slices of the norm of 3D velocity field. The top panel presents the total velocity field, and the second (bottom) panel presents the compressive (solenoidal) component obtained via the Hodge-Helmholtz decomposition. The slices are extracted from the $12\times12\times20$~Mpc box used for the decomposition; the maps are thus $12\times12$~Mpc large, and each pixel is 22.5~kpc large. The slices presented are similar to those in Figs. \ref{left_fil_t_v_div_rot} and \ref{right_fil_t_v_div_rot}, that is at $\Delta x_{\mathrm{cen}}=\{-8,-6,-4,-2,0,2,4,6\}$~Mpc }
            \label{decomp_v}
       \end{minipage}
\end{figure*}

We now study the gas flowing from underdense regions to the filaments, including through a cosmic matter sheet in between for the background filament. We use transverse cuts across at different distances from the cluster centre along the filaments. \\

\subsection{Background filament}

In Fig.\ref{left_fil_t_v_div_rot}, we show four transverse cuts across the background filament at distances $\Delta x_{\mathrm{cen}}=\{-8, -6, -4, -2\}$~Mpc from the cluster centre. The top row presents the temperature, the second the velocity field, the third the divergence, and the bottom row the $x$ component of the vorticity, $\omega_x$. The joint study of these four tracers helps us understand the structuration of this filament and how gas is funnelled towards Virgo. When focusing on the $\Delta x_{\mathrm{cen}}=-8$~Mpc slice, we can see that the filament is embedded in a cosmic matter sheet, as we have already shown in \cite{lebeau2024can}. The gas goes from underdense regions to the matter sheet. It is then accreted by the filament, its typical baryonic radius, which we could define by the maximum of divergence of the velocity field, being only about 2 Mpc at this distance. We can already observe multistreams, heating and stirring in the matter sheet, that divergence separates from underdense regions, and one large eddy around the spine of the filament. \\

Then, in the cut across at $\Delta x_{\mathrm{cen}}=\{-8,-6,-4\}$~Mpc, we can see that the two opposite gas flows fueling the filament are not perfectly aligned, which generate gas coiling around the spine of the filament and thus a large eddy, it is particularly visible on the $\Delta x_{\mathrm{cen}}=-4$~Mpc slice (third column of Fig. \ref{left_fil_t_v_div_rot}). We also observe that when approaching Virgo, the gas in the cosmic sheet and the filament heats up and accelerates. Moreover, if defined with the velocity maximum of divergence, its envelope goes from about 3~Mpc to 4~Mpc, while vortices gain intensity. Finally, when the gas enters the cluster (right column of Fig. \ref{left_fil_t_v_div_rot}), as we have already shown with the longitudinal cut, it is heated and stirred, leading to a large amount of turbulence. We also observe that the gas reaches its highest velocity at $\Delta x_{\mathrm{cen}}=-2$~Mpc and that the divergence defines the cluster baryonic boundary fairly well, which we could relate to the accretion radius. This radius has about the same value as the $R_{\mathrm{vir}}$ but is much smaller than the splashback radius estimated from baryons in \cite{lebeau2024can}. If observable, the divergence of the velocity field might thus be a boundary definition of clusters that could be comparable with $R_{\mathrm{vir}}$. 

\begin{table}[h!]
    \centering
    \caption{Centre of the regions used for the 2D power spectra}
    \renewcommand{\arraystretch}{1.3}
    \begin{adjustbox}{width=0.5\textwidth}
    \begin{tabular}{ c c c c c c c c c }
    \hline \hline
    \multicolumn{9}{c}{Centre of the regions used for the 2D power spectra}\\
    \hline
    $\Delta x_{\mathrm{cen}}$ (Mpc) & -8 & -6 & -4 & -2 & 0 & 2 & 4 & 6 \\
    \hline 
    $\Delta y_{\mathrm{cen}}$ (Mpc) & 3.5 & 1.5 & 1.5 & 0 & 0 & 0 & -1 & -2.5 \\
    \hline 
    $\Delta z_{\mathrm{cen}}$ (Mpc) & 1.5 & 1 & 0.5 & 0 & 0 & 0 & 0 & -0.5 \\
    \hline
    
    \end{tabular}
    \end{adjustbox}
    \label{tab:pos}
    \tablefoot{Each coordinates are given in Mpc with respect to the Virgo's DM halo centre as found by the halo finder.}
\end{table}

\subsection{Foreground filament}

In Fig. \ref{right_fil_t_v_div_rot}, we show three transverse cuts across the foreground filament at distances $\Delta x_{\mathrm{cen}}=\{2,4,6\}$~Mpc from the cluster centre. At larger distances, the foreground filament is much more diluted than the background one. The figure is organised similarly to Fig. \ref{left_fil_t_v_div_rot}. Coherently with what we observed on the longitudinal cut (see Fig. \ref{zoom_right}), we can see that this filament is much more diffuse than the background one. Indeed, at $\Delta x_{\mathrm{cen}}=6$~Mpc, the filament can be identified as the converging point of the velocity field (second row) and the hottest area, but it does not stand out much from its environment. Moreover, there are multiple flows towards it, whereas the background filament is the confluence of the two gas flows from each part of the matter sheet. The gas around this filament directly flows from underdense regions, and we cannot distinguish a cosmic matter sheet in this area. At $\Delta x_{\mathrm{cen}}=4$~Mpc, multiple gas streams are also visible. Still, the filament is more defined as it is hotter, better defined by the divergence, and the gas inside is rotating, as shown by the vorticity. However, it does not have a cylindrical shape, which is certainly due to the multiple streams, which could be seen as smaller-scale filaments \citep[e.g.][]{2010MNRAS.408.2163Aragon-Calvo, 2025MNRAS.539..873Feldbrugge}. Then, at $\Delta x_{\mathrm{cen}}=2$~Mpc, we are roughly at Virgo's virial radius, and the picture is indeed roughly similar to the cut at $\Delta x_{\mathrm{cen}}=-2$~Mpc in the background filament. \\

In conclusion, we can see that the gas swirling starts in this foreground filament, similarly to the background filament, even though it is not as large and collimated. The eddies seem to be transported along the gas flow towards Virgo while becoming more intense and numerous due to shocks and gas accretion at the boundaries of the filament. 

\section{Velocity fields in filaments}
\label{sec:5}

In the previous sections, we presented a qualitative study of the gas flows in the filaments connected to Virgo, showing how eddies are generated and grow towards Virgo. We now decompose the velocity field into its compressive and solenoidal components to quantify their contribution to the total velocity field and characterise turbulence through the 2D total power spectrum. \\

\subsection{Hodge-Helmholtz velocity field decomposition}
\label{sec:5.1}

The velocity field is a mixture of compressive and rotating flows. The Hodge-Helmholtz theorem stipulates that any vector can be separated into a curl-free, a divergence-free, and a harmonic component \citep[e.g.][]{arfken2011mathematical}. In the case of fluid mechanics, the first is the compressive component, the second is the solenoidal component, and we consider the third one to be null in this work, following \cite{valles2021unravelling}. Following \cite{vazza2017turbulence}, we apply this decomposition to the Fourier transform of the velocity field written as $\vec{v}_k=\vec{v}_{k,c}+\vec{v}_{k,s}$  with 

\begin{equation}
     \vec{v}_{k,c}=\vec{k}(\vec{k}\cdot\vec{v}_k)/|k|^2 \quad \textrm{and} \quad \vec{v}_{k,s}=-\vec{k}\times(\vec{k}\times\vec{v}_k)/|k|^2.
\end{equation}

\noindent We can thus efficiently calculate $\vec{v}_{k,c}$ by projecting $\vec{v}_k$ on $\vec{k}$ using Fast Fourier Transform (FFT) \footnote{We used the open code (\url{https://github.com/shixun22/helmholtz}) from Xun Shi }, then return to real space and deduce the solenoidal component as $\vec{v}_s=\vec{v}-\vec{v}_c$. \\

The decomposition has been applied on a rectangular $12\times12\times20$~Mpc box encompassing Virgo and the two filaments. We present slices of this box in Fig. \ref{decomp_v}. The slices are similar to those presented in Figs. \ref{left_fil_t_v_div_rot} and \ref{right_fil_t_v_div_rot}, that is $\Delta x_{\mathrm{cen}}=\{-8,-6,-4,-2,0,2,4,6\}$~Mpc. The top panels show the total velocity field norm, $||\vec{v}||=\sqrt{v_x^2+v_y^2+v_z^2}$, the middle panels show the compressive component norm, $||\vec{v}_c||$, and the bottom panels show the solenoidal component norm, $||\vec{v}_s||$. \\

In every slice and particularly for central slices at $\Delta x_{\mathrm{cen}}=\{-2,0,2\}$~Mpc, we can see that the compressive component (middle row) highlights the boundaries of the baryonic matter sheet, the filaments and the cluster and more generally regions where accretion shocks occur. These features traced by the compressive component are coherent with the ones traced with the divergence of the velocity field presented on the central panels of Figs. \ref{zoom_left} and \ref{zoom_right}, and third rows of Figs. \ref{left_fil_t_v_div_rot} and \ref{right_fil_t_v_div_rot}, which is expected given that the compressive component is, by definition, curl-free. On the other hand, we note that the solenoidal component traces rotation occurring mostly inside large-scale structures. This feature agrees with that traced by the vorticity presented in the right panels of Figs. \ref{zoom_left} and \ref{zoom_right} and bottom rows of Figs. \ref{left_fil_t_v_div_rot} and \ref{right_fil_t_v_div_rot}, which is also expected given that this component is, by definition, divergence-free. To quantify turbulence in these slices, we analyse their power spectrum in the following subsection. 

\begin{figure}

            \centering
            \includegraphics[trim= 20 10 100 0, clip, width=.49\textwidth]{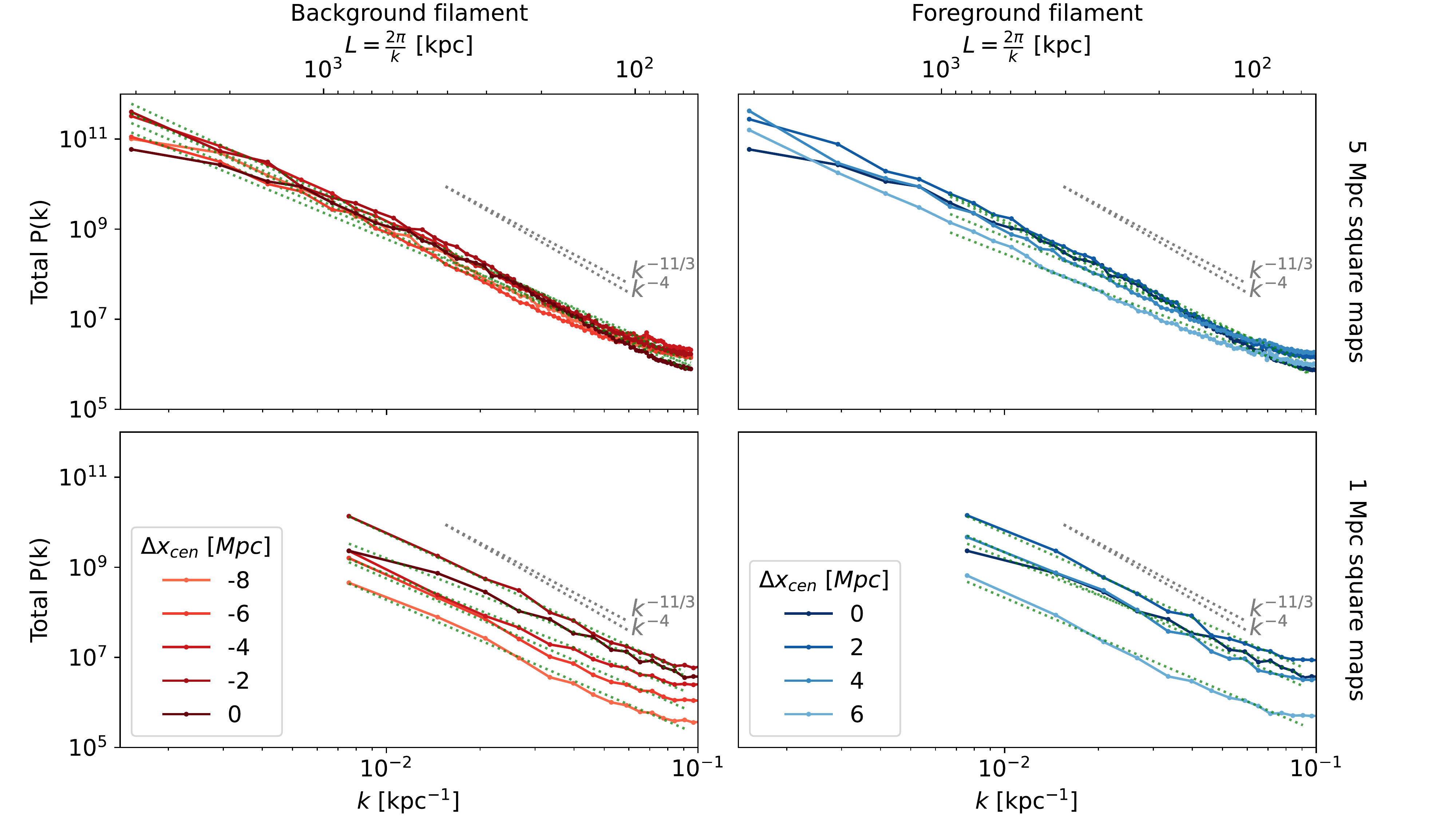}
            \caption{Comparison of the total velocity field power spectra. The top and bottom rows present spectra computed from 5 and 1~Mpc square maps, respectively. The left column presents the total velocity field power spectra in slices in the background filament, that is, from red to black at $\Delta x_{\mathrm{cen}}=\{-8,-6,-4,-2\}$~Mpc, and in Virgo's core for comparison. The right column presents the total velocity field power spectra in slices in the foreground filament, that is, from black to blue at $\Delta x_{\mathrm{cen}}=\{2,4,6\}$~Mpc, and also in Virgo's core for comparison. For comparison, we show the slopes predicted by Kolmogorov's ($P(k)\propto k^{-11/3}$) subsonic turbulence theory and Burgers' ($P(k)\propto k^{-4}$) supersonic turbulence theory.}
            \label{ek_tot_compare}
    
\end{figure}

\begin{table*}[h!]
    \centering
    \caption{Best fit of the total power spectrum slope}
    \renewcommand{\arraystretch}{1.3}
    \resizebox{1\textwidth}{!}{\begin{tabular}{ c c c c c c c c c }
    \hline \hline
    \multicolumn{3}{c}{Kolmogorov : $P(k)\propto k^{-11/3}$ } & \multicolumn{3}{c}{Burgers : $P(k)\propto k^{-4}$ } & \multicolumn{3}{c}{Fit : $P(k)\propto k^{\alpha}$ } \\
    \hline 
    \multicolumn{9}{c}{Best fit of the total power spectrum slope}\\
    \hline
    \hline
    $\Delta x_{\mathrm{cen}}$ (Mpc) & -8 & -6 & -4 & -2 & 0 & 2 & 4 & 6 \\
    \hline 
    $\alpha$ (5~Mpc) & -3.00$\pm$0.03 & -2.89$\pm$0.04 & -3.05$\pm$0.03 & -3.20$\pm$0.03 & -3.18$\pm$0.04 & -3.23$\pm$0.03 & -2.95$\pm$0.03 & -2.85$\pm$0.04 \\
    \hline 
    $\alpha$ (1~Mpc) & -2.99$\pm$0.09 & -3.01$\pm$0.09 & -2.72$\pm$0.08 & -3.19$\pm$0.06 & -2.71$\pm$0.07 & -3.11$\pm$0.08 & -3.10$\pm$0.08 & -2.95$\pm$0.1  \\
    \hline
    
    \end{tabular} 
    }
    \label{tab:fit}
    \tablefoot{The fit is made on the power spectra presented in Fig. \ref{ek_tot_compare} for k<0.095 (L>66~kpc). The first, second and bottom rows present the best fit for the spectra computed from the 5 and 1~Mpc maps, respectively. The standard deviation is given for each fit.}
\end{table*}

\subsection{2D power spectra}

We now quantitatively characterise the velocity field in the slices and compare the contribution of each component to the total velocity field throughout the filaments. To do so, we compute the power spectrum, $P(\vec{k})$, that we define for the total velocity field as 

\begin{equation}
    P(\vec{k}) = \frac{1}{2}|\vec{\tilde{v}}(\vec{k})|^2=\frac{1}{2}(|\tilde{v}_x(\vec{k})|^2 + |\tilde{v}_y(\vec{k})|^2 + |\tilde{v}_z(\vec{k})|^2),
    \label{equ:P(k)}
\end{equation}

\noindent with $\vec{\tilde{v}}(\vec{k})$ the Fourier transform of the total velocity field. The formula is similar for the compressive and solenoidal components of the velocity field. It is worth noting that we compute the power spectrum in the rest frame of Virgo's centre; thus, we do not subtract the coherent flow, in other words, the average gas velocity in each slice, in the filaments from the total velocity, as done in, e.g. \citet{vazza2017turbulence}. Tests have shown that, thanks to the properties of the Fourier Transform, it yields exactly the same results. We thus compare the contribution of each component to the total velocity field and do not use the terms "solenoidal turbulence" and "compressive turbulence" as in \citet{vazza2017turbulence}. We only talk about turbulence when discussing the total power spectrum.\\

To investigate the velocity field and its components in the core of the filaments and the cluster, but also in their direct surrounding, we compute the 2D power spectrum on 1 and 5~Mpc square maps centred on the core of the filament. The selected regions are displayed as white squares on the top row of Fig. \ref{decomp_v} and their centre can be found in Tab. \ref{tab:pos}. Since our study focuses on the gas flows, we defined manually the centre of the selected regions as the point of convergence of the velocity flows, as shown in the second row of Figs. \ref{left_fil_t_v_div_rot} and \ref{right_fil_t_v_div_rot}. We thus purposefully did not use any filament finders, which usually rely on the positions of DM particles or galaxies. Using a filament finder to define the centre of the maps as the spine of the filament could give a slightly different position compared to the hand-selected centres of the slices, which could marginally modify the velocity power spectrum of the 1~Mpc large maps, but should not modify the velocity power spectra computed from larger maps. The 2D power spectrum is calculated similarly to the 3D case \footnote{Adapting the 2D Power spectrum routine of the Pylians library \citep{Pylians}} (see Equ. \ref{equ:P(k)}), but only on one plane extracted from the 3D regular grid after the Hodge-Helmoltz decomposition, as if computed on a map but without projecting the components of the velocity field along a sightline as in \citet{lebeau2025_vel_core}.   \\

\begin{figure}

            \centering
            \includegraphics[trim = 20 10 80 0,clip, width=.49\textwidth]{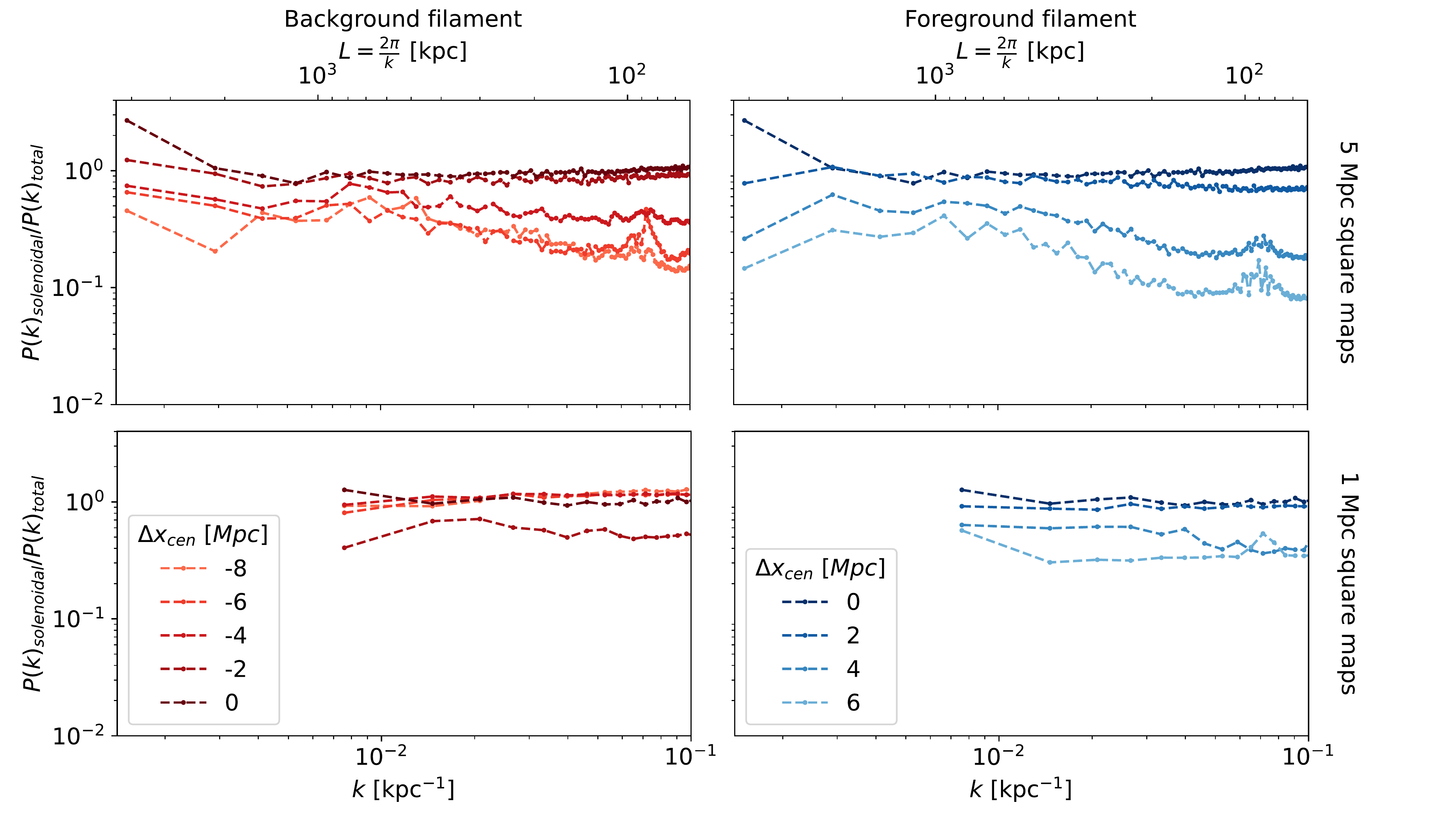}
            \caption{Ratio of the power spectrum of the solenoidal component over the total velocity field. The top and bottom rows present ratios of spectra computed from, respectively, 5 and 1~Mpc square maps. The left column presents the ratios in slices in the background filament, that is, from red to black at $\Delta x_{\mathrm{cen}}=\{-8,-6,-4,-2\}$~Mpc, and in Virgo's core for comparison. The right column presents the ratios in slices in the foreground filament, from black to blue at $\Delta x_{\mathrm{cen}}=\{2,4,6\}$~Mpc, and Virgo's core for comparison.}
            \label{ek_ratio_st}
    
\end{figure}

\begin{figure}

            \centering
            \includegraphics[trim = 20 10 100 0,clip, width=.49\textwidth]{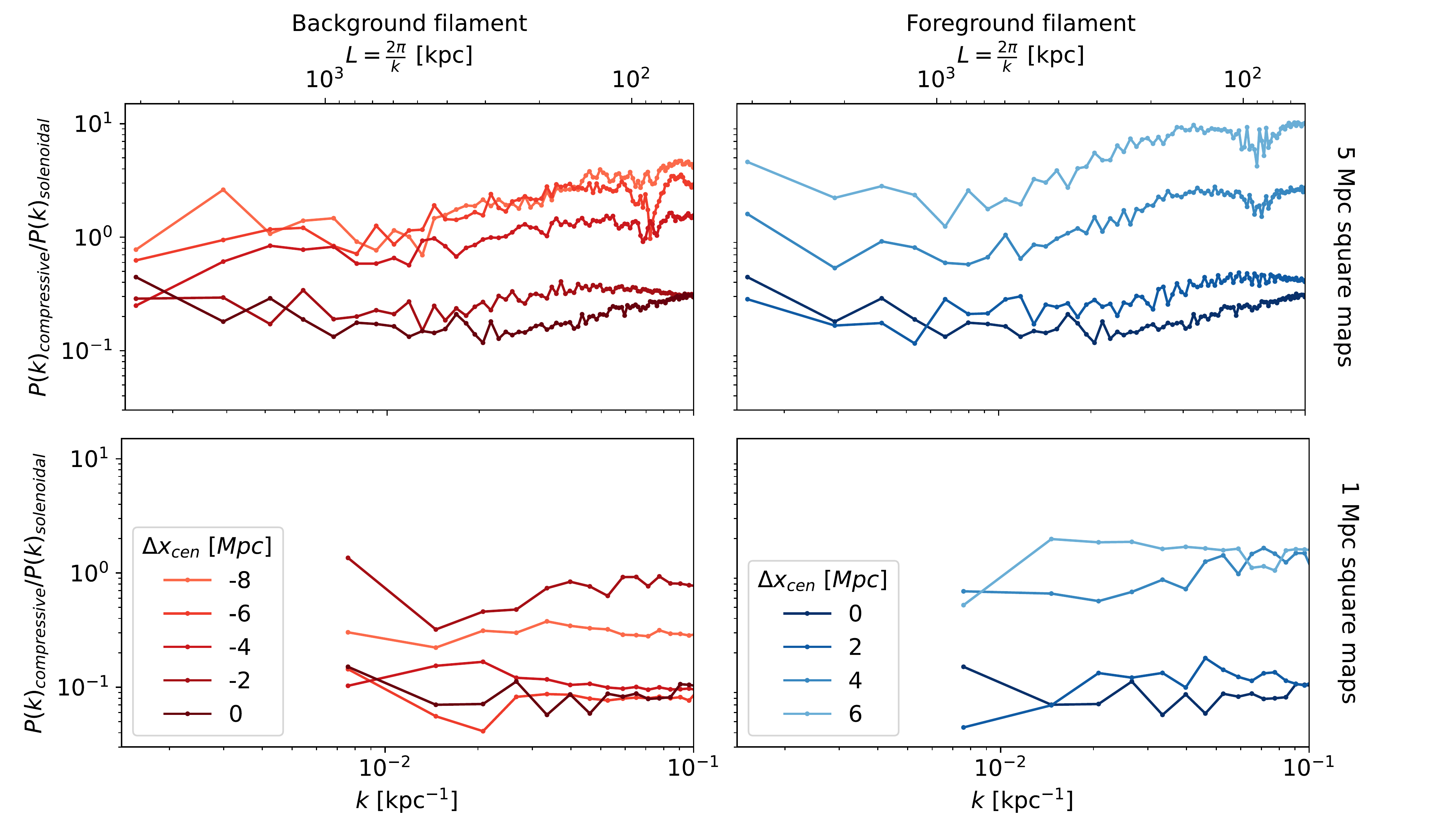}
            \caption{Ratio of the power spectrum of the compressive over the solenoidal component of the velocity field. The top and bottom rows present ratios of spectra computed from, respectively, 5 and 1~Mpc square maps. The left column presents the ratios in slices in the background filament, that is, from red to black at $\Delta x_{\mathrm{cen}}=\{-8,-6,-4,-2\}$~Mpc, and in Virgo's core for comparison. The right column presents the ratios in slices in the foreground filament, from black to blue at $\Delta x_{\mathrm{cen}}=\{2,4,6\}$~Mpc, and Virgo's core for comparison.}
            \label{ek_ratio_cs}
    
\end{figure}

For the sake of clarity, we only analyse here the power spectra and their ratios computed from 1 and 5~Mpc square maps; all the 2D spectra, including those computed from 2~Mpc square maps, can be found in Fig. \ref{ek_2D_all} in Appendix \ref{app:2D_ek_all}. We chose a 2D power spectrum on slices over 3D to avoid overlap, and thus possible correlations, between studied regions given that the slices are only 2~Mpc apart; we discuss this choice in Sect. \ref{sec:6}. To simplify the analysis and enable comparisons of the spectra, we only show the 2D power spectra of the total velocity field in Fig. \ref{ek_tot_compare}, the solenoidal-to-total power spectra ratios in Fig. \ref{ek_ratio_st} and the compressive-to-solenoidal power spectra ratios in Fig. \ref{ek_ratio_cs}. The same figures for the spectra computed from 2~Mpc square maps can be found in Fig.\ref{app:ek comp 2Mpc} in Appendix \ref{app:ek comp 2Mpc}. In these three figures, the top and bottom panels present the spectra, or ratios, computed from 5 and 1~Mpc square maps, respectively. The left column presents the total power spectrum, or ratios, in the background filament at distances $\Delta x_{\mathrm{cen}}=\{-8,-6,-4,-2,0\}~\mathrm{Mpc}$ from red to black and the right column present those in the foreground filament at distances $\Delta x_{\mathrm{cen}}=\{0,2,4,6\}~\mathrm{Mpc}$ from black to blue.\\

First, we can visually notice, in Fig. \ref{ek_tot_compare}, that all the total power spectra have a negative slope, typical of a turbulent cascade \citep{richardson1922weather}; it is also the case for each component with relatively similar slopes compared to their associated total power spectrum (see Fig. \ref{ek_2D_all} in Appendix \ref{app:2D_ek_all}). We can also see that, for both filaments and independently of the map size, although less visible for the 5~Mpc square maps, there is an amplitude increase when approaching Virgo, meaning more intense eddies due to more energy injection by shocks at multiple scales, as shown by the divergence of the velocity field in the third panel of Figs. \ref{left_fil_t_v_div_rot} and \ref{right_fil_t_v_div_rot}. However, in Virgo's core, the amplitude is lower than at $\Delta x_{\mathrm{cen}}=\{-2,2\}~\mathrm{Mpc}$, because kinetic energy is dissipated and converted into heat and so the eddies are slowed down when entering the cluster because of the termination shock at $R_{\mathrm{vir}}\approx$ 2~Mpc. \\

We now compare the power spectra slopes. For simplicity, we only discuss the total velocity and do not compare the slopes of each component, given that they have similar behaviours. To do so, we fitted the power spectra to a power law $P(k)=Ak^{\alpha}$, the best-fit values for the slope $\alpha$ can be found in Tab. \ref{tab:fit}. The slopes were fitted for $k<0.095$ ($L$>66~kpc); which is chosen to be slightly smaller in $k$ than the Nyquist frequency, $f_N$, which is twice the best cell resolution of 22.5~kpc, to avoid contamination in the fit from numerical flattening close to $f_N$ (see Fig. \ref{fig:app:nyq freq}). Each row represents a given square map size from which the power spectra were computed. We observe that in the range $\Delta x_{\mathrm{cen}}=[-8,-4]$~Mpc for the background filament, whatever the map size, the slope does not vary. It gets steeper at $\Delta x_{\mathrm{cen}}=-2$~Mpc, likely due to the termination shock front and the associated small-scale shocks within $R_{vir}$ transferring more kinetic energy into heat. Then, in the foreground filament, the slope gets smoothly steeper when approaching Virgo and is the steepest at $\Delta x_{\mathrm{cen}}=2$~Mpc, most likely for the same reason as in the background filament. In Virgo's core, the slope is less steep than at $\Delta x_{\mathrm{cen}}=-2$ and 2~Mpc, particularly on the spectra computed from the 1~Mpc square map. However, it is worth noting that visually, we observe a steeper slope at the smallest scales in the core for the power spectrum computed from the 5~Mpc square map (black line on the top left panel of Fig. \ref{ek_tot_compare}), and a small flattening at the largest scales, which could be the injection scale, but since we do not compute the power spectra on larger scales we cannot confirm this speculation. When restricting the fitting interval to k=[0.006,0.095]~$\mathrm{kpc^{-1}}$, we still observe an increase in slope along the filament, but the steepest slope is in Virgo's core. It is certainly due to the termination shock at $R_{\mathrm{vir}}$ that is covered in the 5~Mpc square map. \\

We then discuss the contribution of each component to the total velocity field by computing the ratios of power spectra in the slices, from Virgo's core to far into the filaments. In Virgo's core, $\Delta x_{\mathrm{cen}}=0$~Mpc is displayed in black in Figs. \ref{ek_ratio_st} and \ref{ek_ratio_cs}, whatever the map size, the solenoidal to total ratio is almost flat and equal to one, meaning that the solenoidal component largely dominates the velocity field and that there is indeed almost no contribution to the overall velocity from the compressive component. Moreover, the compressive to the solenoidal ratio in Fig. \ref{ek_ratio_cs} is generally the lowest among the slices, whatever the map size. It is in agreement with \citet{vazza2017turbulence} who also found a solenoidal-dominated turbulence in the core of their simulated cluster. \\

At 2~Mpc from Virgo's core, $\Delta x_{\mathrm{cen}}=-2$ and $2$~Mpc, that is, at Virgo's connection with both the background and foreground filaments and at about the virial radius, the solenoidal component contains most of the energy in the flow of gas again. Still, the energy of the compressive component is slightly higher than in Virgo's core, as we see in Fig. \ref{ek_ratio_cs}, whatever the map size. In particular, for the 1~Mpc square map in the background filament (bottom left panels of Figs. \ref{ek_ratio_st} and \ref{ek_ratio_cs}), the solenoidal to total ratio is the lowest, and the solenoidal component spectrum is almost equal to that of the compressive component at small scales. It might be because the compressive component traces the gravitational accretion around the spine of the filament.  \\ 

When focusing on the background filament, $\Delta x_{\mathrm{cen}}=\{-8,-6,-4,-2\}$~Mpc, the amplitude ratio between the spectra of the two components depends on the map size, as we see in the left column of Fig. \ref{ek_ratio_cs}. Looking at the spectrum computed from the 1~Mpc square maps (bottom row), we observe that the solenoidal component dominates the compressive one, like in Virgo's core. It is due to the large eddy in the core of the filament (see second row of Fig. \ref{left_fil_t_v_div_rot}, particularly the second panel at $\Delta x_{\mathrm{cen}}=-6$~Mpc). However, the spectrum computed from the 5~Mpc square size maps shows a different trend; the compressive component dominates over the solenoidal at $\Delta x_{\mathrm{cen}}=-8$~Mpc, and we observe a smooth progressive inversion as the ratio goes from between 1 and 5 at $\Delta x_{\mathrm{cen}}=-8$~Mpc to around 0.2 at $\Delta x_{\mathrm{cen}}=-2$~Mpc. It indicates that, in this large and collimated filament, the gas accretion, highlighted by the compressive component of the velocity field, contributes to the generation of rotation far away and is then superseded by the contribution of large-scale eddies in the filament as the latter grows in radius. In addition, the slope of the solenoidal component is steeper than that of the compressive component on a 5~Mpc square size map at $\Delta x_{\mathrm{cen}}=-8$~Mpc. It is because at this distance there is not much rotation at small scales, only one large eddy, see the vorticity on transverse slices in Fig. \ref{left_fil_t_v_div_rot}. Most of the power comes from the laminar flows accreting matter that the 5~Mpc square size map encompasses. Consequently, the power spectrum of the solenoidal component shows power at large scales and then drops rapidly. We observe the same feature in the foreground filament on the 5~Mpc square size map at $\Delta x_{\mathrm{cen}}=6$~Mpc.\\

Now focusing on the foreground filament ($\Delta x_{\mathrm{cen}}=\{2,4,6\}$~Mpc, right column of Fig. \ref{ek_ratio_cs}), we observe that, whatever the map size, the solenoidal component dominates over the compressive one at $\Delta x_{\mathrm{cen}}=2$~Mpc. Then, the two components have more or less equivalent amplitude at $\Delta x_{\mathrm{cen}}=4$~Mpc, and the compressive component dominates over the solenoidal one at $\Delta x_{\mathrm{cen}}=6$~Mpc. In particular, at this distance, the ratio of compressive to solenoidal is of the order of 5 to 10 from the 5~Mpc square maps and 1 to 2 from 1~Mpc square maps. It shows that there is indeed a higher contribution from the compressive component further away from the core of the filament due to gas accretion, as also visible in the right column of Fig. \ref{right_fil_t_v_div_rot}. However, in the right column of Fig. \ref{ek_ratio_cs}, we see that, whatever the map size, there is a continuous decrease of the compressive-to-solenoidal ratio while approaching Virgo. It shows that, in the case of a less collimated filament with multi-stream accretion, there is no obvious eddy around the spine of the filament. The cosmic gas looks to be funnelled towards Virgo, thus generating turbulence in a less structured manner than in the well-collimated background filament, where the gas is accreted from two streams coming from both sides of the matter sheet. \\

Overall, the conjugate increase in amplitude and steepening of the slope of the velocity power spectrum up to Virgo's outskirts shows that more and more energy is injected into the turbulent cascade because of shocks, which induce more conversion of kinetic energy into heat. In addition, the velocity field transits from a compressive- to solenoidal-dominated regime. All these findings tend to show that turbulence in the filaments develops towards the cluster, finally transiting at the termination shock from a compressible supersonic state, as modelled by \citet{burgers1948mathematical}, in filaments to a more developed subsonic and solenoidal-dominated turbulence in Virgo, close to the ideal fully developed, subsonic and incompressible case of \citet{kolmogorov1941local}. We compare our findings to the aforementioned models in the following Sect. \ref{sec:6}. \\

\section{Discussion}
\label{sec:6}

In the following discussion, we compare this work to pioneering ones on gas temperature and velocity in large-scale structures, discuss the impact of computing 2D power spectra instead of 3D and propose interpretations of the results. Finally, we propose future extensions to this work given studies on turbulence in galaxy clusters. \\

The search for the origin of the acceleration of cosmic rays in the cosmic web and the thermalisation of cosmic baryons led to the study of shocks during large-scale structure formation. \cite{ryu2003waves} and \cite{pfrommer2006shock} produced cosmological simulations in which they both found signatures of accretion shocks at the boundaries of cosmic sheets, filaments and clusters. These so-called external shocks \citep{ryu2003waves}, with a Mach number at shocks estimated from the gas temperature jump following the Rankine-Hugoniot conditions, were up to $M=100$ or higher. They also found internal shocks in filaments and haloes with a Mach number around $M=2$. However, they did not extend their studies to the eddies generated by these shocks. \\

\cite{ryu2008turb}, on the contrary, studied the cascade of vorticity to estimate the generated magnetic field strength. In an MHD simulation reproducing the conditions in the ICM, \cite{porter2015vorticity} tested the generation of turbulence from solenoidal-only and compressive-only forcing; they have shown that both these forcing processes led to a mixture of solenoidal and compressive turbulent motions; in particular, compressive flows generate shocks which induce vorticity. Our results generally agree with theirs as we observe the contribution of both compressive and solenoidal components to the velocity field. However, our Virgo replica is a hydrodynamical-only simulation, thus not allowing a detailed comparison given the correlation between magnetic fields and turbulence in their work. \\

In a series of works, turbulence has been simulated in the Inter-Galactic Medium (IGM) \citep{iapichino2011turbulence} and clusters' outskirts \citep{2017MNRAS.469.3641I_Iapichino}, both using the AMR code ENZO \citep{2014ApJS..211...19B_Bryan} and including a subgrid model for unresolved turbulence, computing the energy content on unresolved length-scales. In \cite{iapichino2011turbulence}, they have shown that turbulence is mostly generated by shocks in the WHIM and merger-induced shear flows in the ICM, which agrees with our work. In \cite{2017MNRAS.469.3641I_Iapichino}, they particularly studied the shocked ICM in cluster outskirts, focusing on the energy budget, which, in their case, is dominated by the major merger. This both heats the gas and induces turbulence, which we also observed at $\Delta x_{\mathrm{cen}}=\{-2,2\}$~Mpc, although we are observing the shocked gas accreted from filaments and not the consequences of a merger. \\

Recently, \citet{2024MNRAS.52711256L_Lu} studied filaments connecting massive high-redshift galaxies following a similar tomographic approach. In a zoom-in simulation encompassing three $\sim$ 1~Mpc filaments in-between $\sim 10^{12}\mathrm{M_{\odot}}$ haloes at z $\sim$ 4, they have shown that their filament could be separated in an outer "thermal", intermediate "vortex" and inner "stream" zones, the "stream" zone being the cold gas stream feeding galaxies. Although the methodology is similar to ours, we find very different results; in particular, we observe a large eddy around the spine of the background filament. It comes from the fact that we study much hotter, longer and thicker filaments connected to a $\sim 4 \times 10^{14}\mathrm{M_{\odot}}$ galaxy cluster at z=0. \\

We now compare our results with theoretical predictions. Some studies (e.g. \citeauthor{vazza2017turbulence} \citeyear{vazza2017turbulence}) focused on the 3D power spectrum, allowing for a direct comparison. However, in our work, we computed 2D power spectra; we thus need to account for the slope difference between 2D and 3D power spectra. Other works, e.g. \citet{2012MNRAS.422.2712Zhuravleva} in mock X-ray observations, predicted a proportionality relation, thus a similar slope, between 2D and 3D power spectra. However, in our study, since we compute the velocity power spectrum on a slice and not a projection, we expect that $P_{3D}(k)k^3\sim P_{2D}(k)k^2$, which rewrites as $P_{3D}(k)/P_{2D}(k)\sim k^{-1}$. To verify this, we computed the 3D power spectrum in a 1~Mpc side cube in the cluster core region and compared it with its 2D counterpart. Results are presented in Fig. \ref{2D vs 3D test}. We also compared the spectra in the other slices, which can be found in Appendix \ref{app:2D vs 3D spectra}. We did not compute the 3D power spectra on larger cubes to avoid overlap and possible correlations, as we discussed above.\\

First, by fitting the slope of the power spectra as $P(k)=Ak^{\alpha}$, we have $P_{3D}(k)/P_{2D}(k)\propto k^{\alpha_{3D}-\alpha_{2D}}$. Computing the difference between the 3D slope and the 2D slope in each case (see Table \ref{tab:app alphas diff}), we can observe that the difference is always close to $-1$, which agrees well with the theoretical prediction. Then, we can observe that the 3D power spectrum in the core of Virgo is in excellent agreement with subsonic \citet{kolmogorov1941local} turbulence theory, predicting a $-11/3\approx-3.67$ slope. Moreover, the 3D power spectrum computed around the spine of the filaments is in quite good agreement with the supersonic Burgers' \citep{burgers1948mathematical} turbulence theory predicting a $-4$ slope (see Fig. \ref{app:2D vs 3D spectra}). It seems to show that, at least in 1~Mpc cube boxes in the core of Virgo and around the spine of the filaments, turbulence has reached the stationary regime. However, it does not imply that turbulence is in a stationary regime within $R_{vir}\sim2$~Mpc, given the transition from supersonic to subsonic state at this distance. In the filaments, stationary supersonic turbulence might be maintained by the steady accreting flows of matter from the outskirts of the filaments, themselves fed by flows arriving from a cosmic sheet of matter and local voids. \\

\begin{figure}
    \centering
    \includegraphics[width=0.99\linewidth]{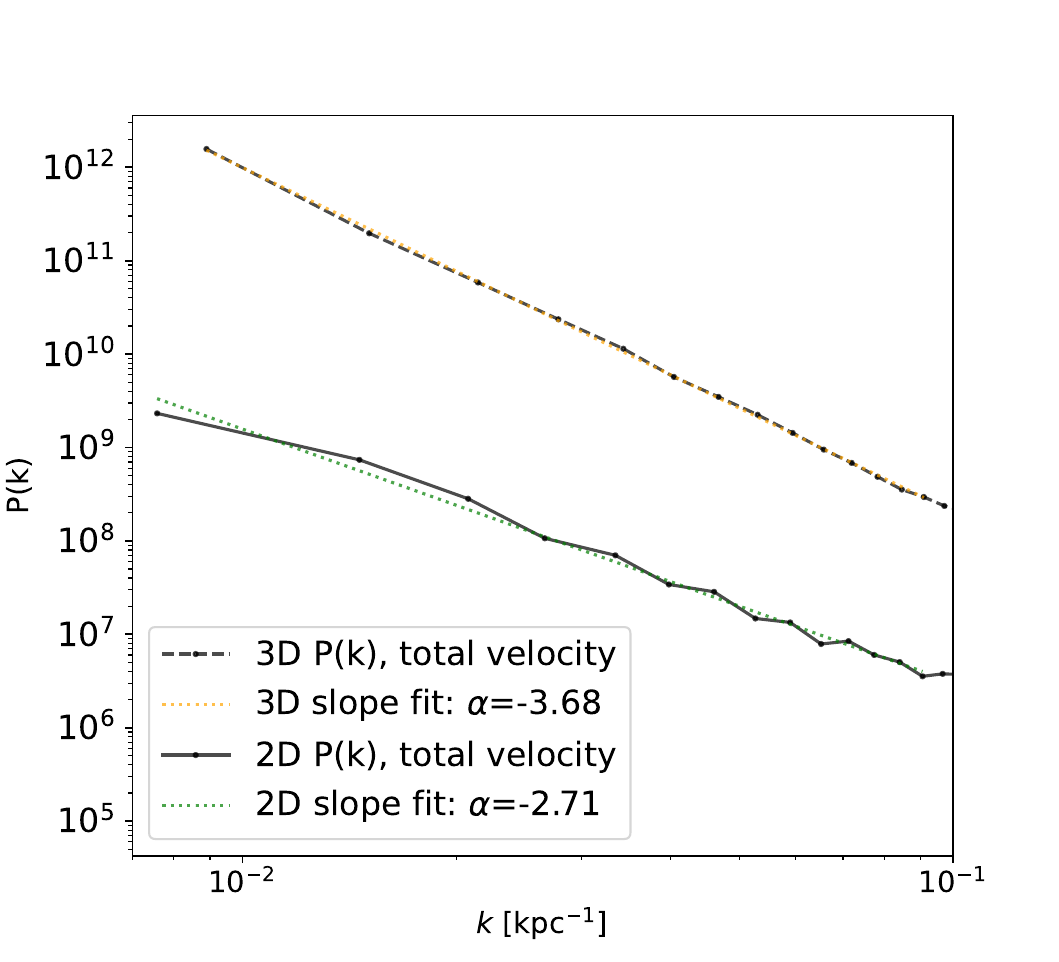}
    \caption{Comparison between the 2D (black solid line) and 3D velocity power spectra (dashed black line), respectively computed from a 1~Mpc square map and a 1 Mpc side cube, both centred on the core of Virgo. The best-fit value of the slope is displayed in dotted coloured lines.}
    \label{2D vs 3D test}
\end{figure}

Finally, we could better characterise turbulence in filaments by calculating the Froude number and other stratified turbulence parameters such as the Brunt–Väisälä (BV) frequency (e.g. \citeauthor{mohapatra2020turbulence} \citeyear{mohapatra2020turbulence}), conduct a more thorough study of the velocity field by analysing the enstrophy \citep[e.g.][]{vazza2017turbulence,valles2021troubled} or use continuous wavelets transform \citep[e.g.][]{2024ApJ...974..107Wang}. We plan to conduct such a study in future work where we will also extend our analysis to a larger number of filaments, the Virgo replica and its filaments not being representative of the cluster population as shown in \cite{sorce2021hydrodynamical}, and study the time evolution of turbulence in filaments through multiple simulation snapshots \citep[e.g.][]{wang2023turbulent} when we only studied the Virgo replica at redshift $z=0$. This will be done with the upcoming LOCALIZATION\footnote{https://localization.ias.universite-paris-saclay.fr/} cosmological simulation reproducing the Local Universe. 

\section{Conclusion}
\label{sec:7}

In this work, we studied velocity fields in filaments connected to the replicated Virgo cluster by investigating longitudinal and transverse slices, first by visual inspection and then by decomposing the velocity field and computing the power spectra. \\

In addition, we compared this work with previous studies on the generation of turbulence in the large-scale structures of the Universe, showing a rather good agreement. Still, given the differences in size, resolution and methods of these works, we could not conduct detailed comparisons. We also discussed comparing the slopes of 2D power spectra to 3D power spectra and theoretical predictions. We have shown that we have a difference in slope between our 3D and 2D velocity power spectra, which seems to agree well with theoretical expectations. We found that the slopes of the 3D power spectra are in excellent agreement with Kolmogorov's subsonic turbulence theory in the core of Virgo and in good agreement with Burgers' supersonic turbulence theory around the spine of the filaments, most likely showing that turbulence reached the stationary regime in these regions. We concluded by proposing improvements and extensions to this work, which we could conduct in the near future. \\

On the one hand, the study of the longitudinal slice shows how cosmic gas streams in filaments towards Virgo. On the other hand, the transverse slices investigation reveals the inner structure of the filaments and how gas flows from underdense regions to matter sheets and then filaments. In particular, the background filament is long, well-structured, and connected to a large matter sheet, whereas the foreground one is much smaller, more diffuse and funnels the gas towards Virgo. Despite these differences, gas swirls inside both, as shown through the vorticity. Accretion shocks at the filament's boundary mostly generate this swirling, which we traced through the divergence of the velocity field. Moreover, the gas accelerates when approaching Virgo and is shocked and heated once again when entering it, roughly at the virial radius, thus accelerating gas swirling. \\

By computing the 2D total power spectrum, we found an increase in amplitude and steepness towards Virgo in both filaments, with the maxima at $\Delta x_{\mathrm{cen}}=$-2 and 2~Mpc, that is at roughly the virial radius, meaning more and more kinetic energy injected into turbulence due to shocks, which is then partly transferred into smaller eddies in the cascade and partly dissipated into heat. In Virgo's core, the power spectrum amplitude is lower than at the virial radius, and the slope is less steep. This tends to show that turbulence reaches a more developed state as it approaches the cluster, and transits from a supersonic and compressible state in the filaments to a more developed subsonic and almost incompressible state in the ICM, close to the ideal model of \citet{kolmogorov1941local}.\\

Then, we decomposed the velocity field into compressive and solenoidal components and studied the compressive-to-solenoidal and solenoidal-to-total power spectra ratios. We found a transition from a compressive-dominated to a solenoidal-dominated velocity field when approaching Virgo for both filaments, reinforcing the scenario of transition from Burgers-like to Kolmogorov-like turbulence. Moreover, the power spectra extracted from the square maps highlighted the different turbulent scales and eddies in those two filaments. In the background filament, the spectra computed from the 1~Mpc square maps show that turbulence is dominated by one large eddy in the core of the filament, and the spectra computed from the 5~Mpc square maps show the swirling generation by gas accretion from the cosmic matter sheet. In the foreground filament, on the contrary, the power spectra show the same smooth compressive-to-solenoidal-dominated transition regardless of the map size. \\

To conclude, the case study of turbulence in Virgo's filaments, the most advanced to date to our knowledge, thanks to the resolution of the simulation, shows how accretion shocks and instabilities generate gas rotation in cosmic filaments and clusters and that the velocity field goes from a compressive- to a solenoidal-dominated regime. It demonstrates that the velocity field in filaments is a precursor of that in clusters. It paves the way for future observations of gas dynamics in the core of the clusters, and potentially in outskirts or bridges, in the X-ray with XRISM or Athena, or even in filaments via 21cm HI with LOFAR or SKA.

\begin{acknowledgements}
The authors thank the referee for their helpful comments to improve this article. The authors acknowledge the Gauss Centre for Supercomputing e.V. (www.gauss-centre.eu) for providing computing time on the GCS Supercomputers SuperMUC at LRZ Munich. This work has been supported by the grant agreements ANR-21-CE31-0019 / 490702358 from the French Agence Nationale de la Recherche / DFG for the LOCALIZATION project. This work has been supported as part of France 2030 program ANR-11-IDEX-0003. The authors thank Florent Renaud for sharing the {\tt rdramses} {\tt RAMSES} data reduction code. The authors thank Eugene Churazov for his precious comments on the velocity power spectrum discussion. SZ thanks Philipp Busch for the early discussions on this topic and acknowledges the hospitality of the Institute d'Astrophysique Spatiale, at the initial phases of this project. SZ also acknowledges support from the Israel Science Foundation grant No. 1388/24.
   
\end{acknowledgements}

\bibliographystyle{aa} 
\bibliography{bibliography}

\appendix

\onecolumn

\section{3D visualisation of the slices}
\label{appendix 3D vis}

\begin{figure*}[h!]
        \begin{minipage}[s]{1\textwidth}
            \centering
            \includegraphics[trim= 20 30 100 50 ,width=.85\textwidth]{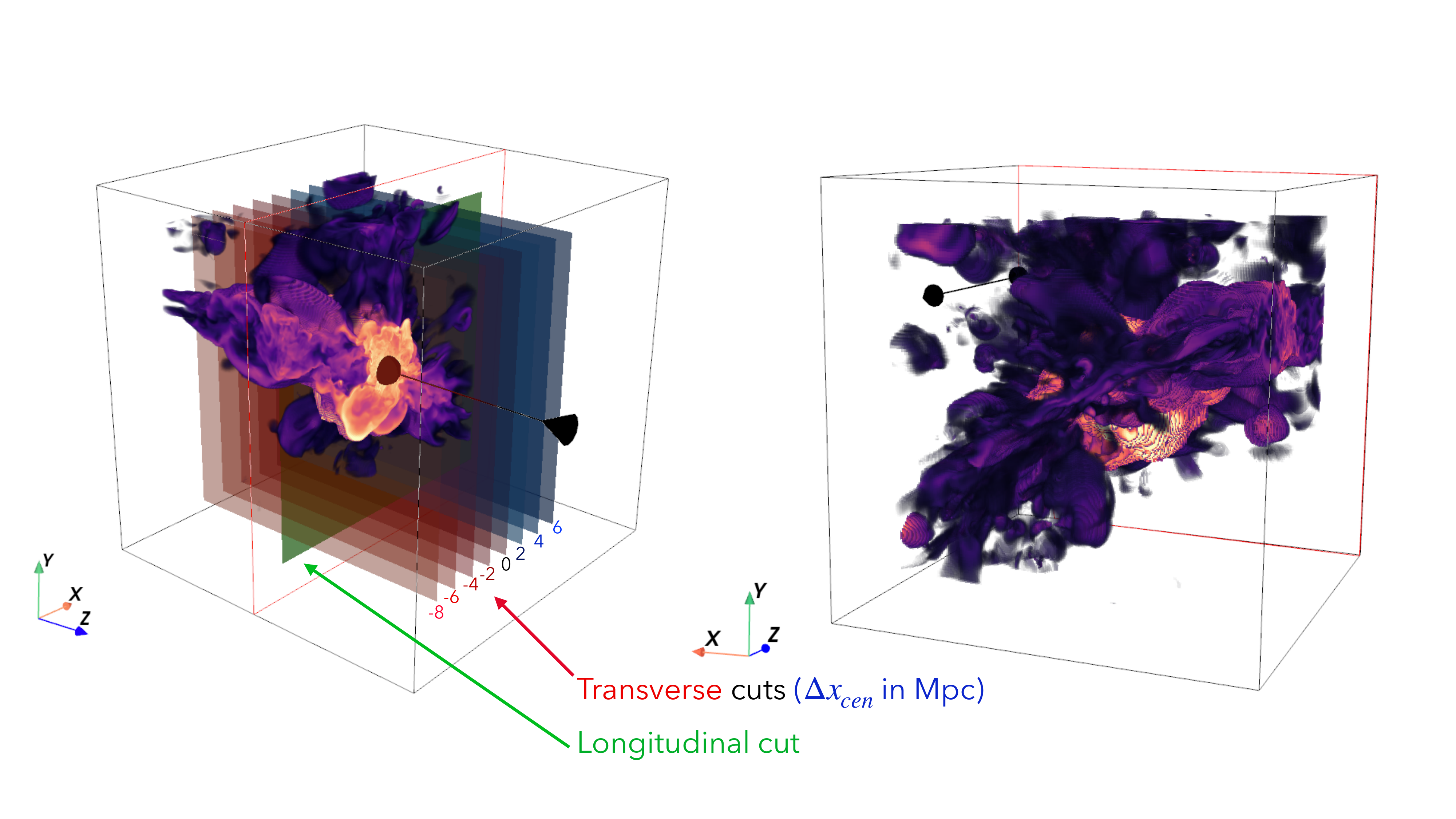}
            \caption{3D visualisation of the temperature in Virgo and its filaments. The green plane represents the longitudinal cut presented in Figs. \ref{long_cut},\ref{zoom_left} and \ref{zoom_right}, similarly to the slice shown on the left panel. The red to black to blue planes represent the transverse cuts presented in Figs. \ref{left_fil_t_v_div_rot} and \ref{right_fil_t_v_div_rot}. On the right panel, the sheet of matter embedding the filaments is in the bottom-left to top-right diagonal. This clipping and slicing visualisation was made using the PyVista Python library \citep{sullivan2019pyvista}. A video showing the simulation at different angles is available at this link: \url{https://youtu.be/VZwacpKZNH4}.}
            \label{app:3D_vis}
       \end{minipage}
\end{figure*}

\section{Electron density in transverse slices}
\label{appendix ne transverse}

\begin{figure*}[h!]
        \begin{minipage}[s]{1\textwidth}
            \centering
            \includegraphics[trim= 20 0 20 10, width=.82\textwidth]{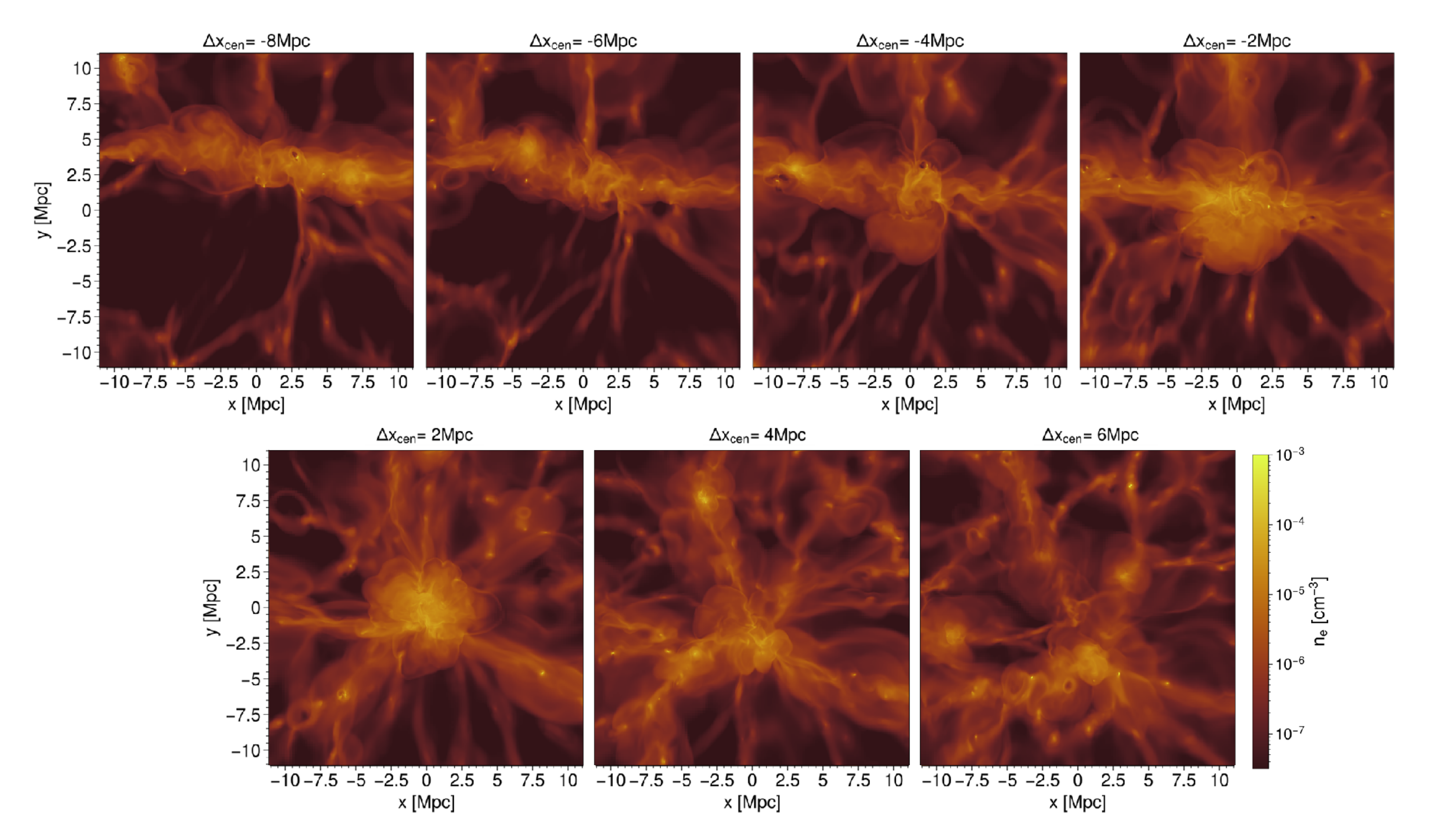}
            \caption{Transverse slices of the electron density along the background filament (top) and foreground (bottom) filaments. On the top row, from left to right, we present cuts at $\Delta x_{cen}=\{-8,-6,-4,-2\}$~Mpc. On the bottom row, from left to right, we present cuts at $\Delta x_{cen}=\{2,4,6\}$~Mpc. Similarly to longitudinal cuts, the maps are 22.122 Mpc wide, contain $\mathrm{15728^2}$ pixels, and are centred on Virgo's centre.}
            \label{app:ne transervse}
       \end{minipage}
\end{figure*}

\newpage

\section{Additional quantities in the zoomed regions}
\label{appendix zoom}

\begin{figure*}[h!]
        \centering
        \begin{minipage}[s]{1\textwidth}
            \centering
            \includegraphics[width=.9\textwidth]{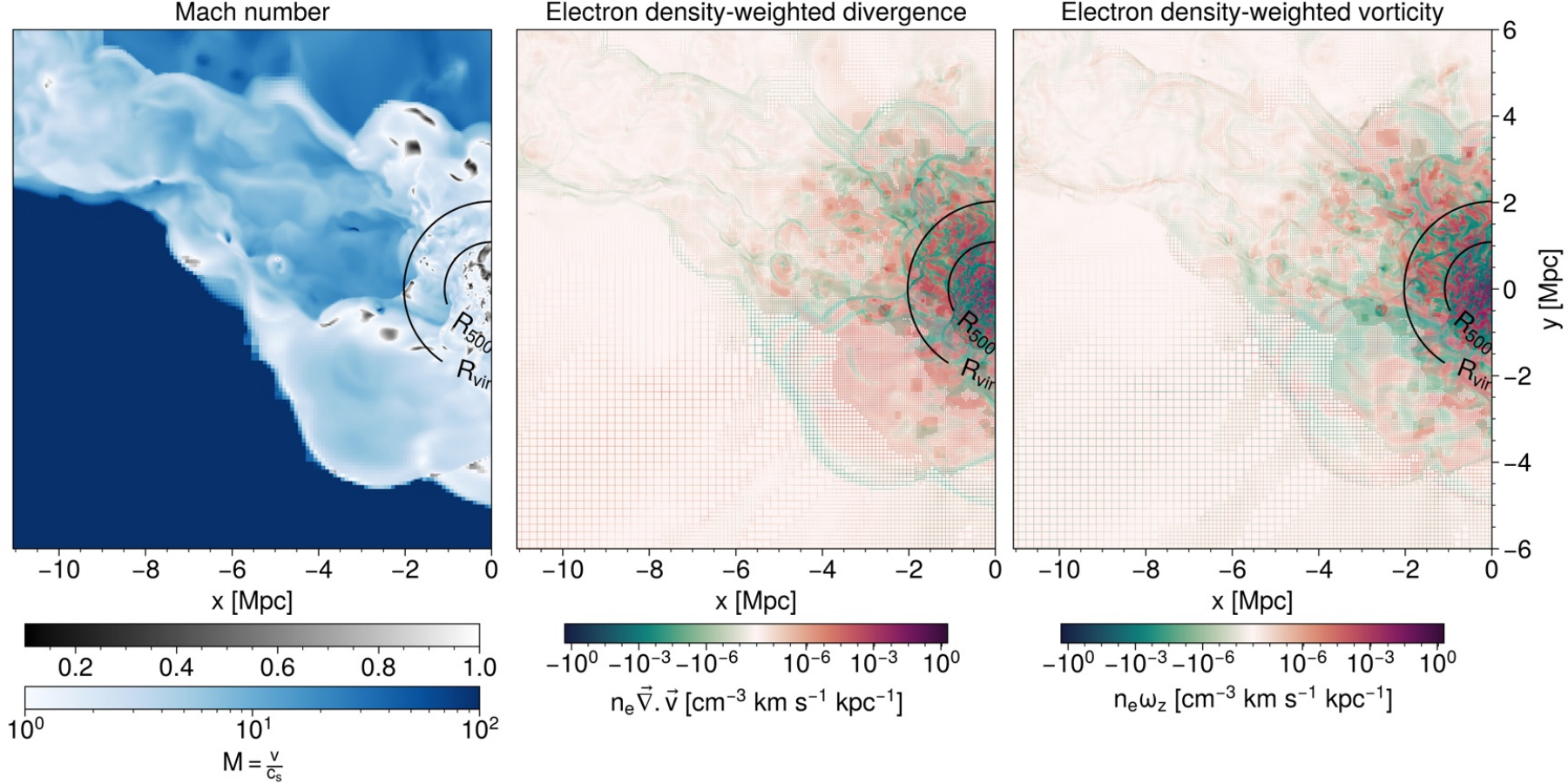}
            \hspace{0.1cm}
            \includegraphics[width=.9\textwidth]{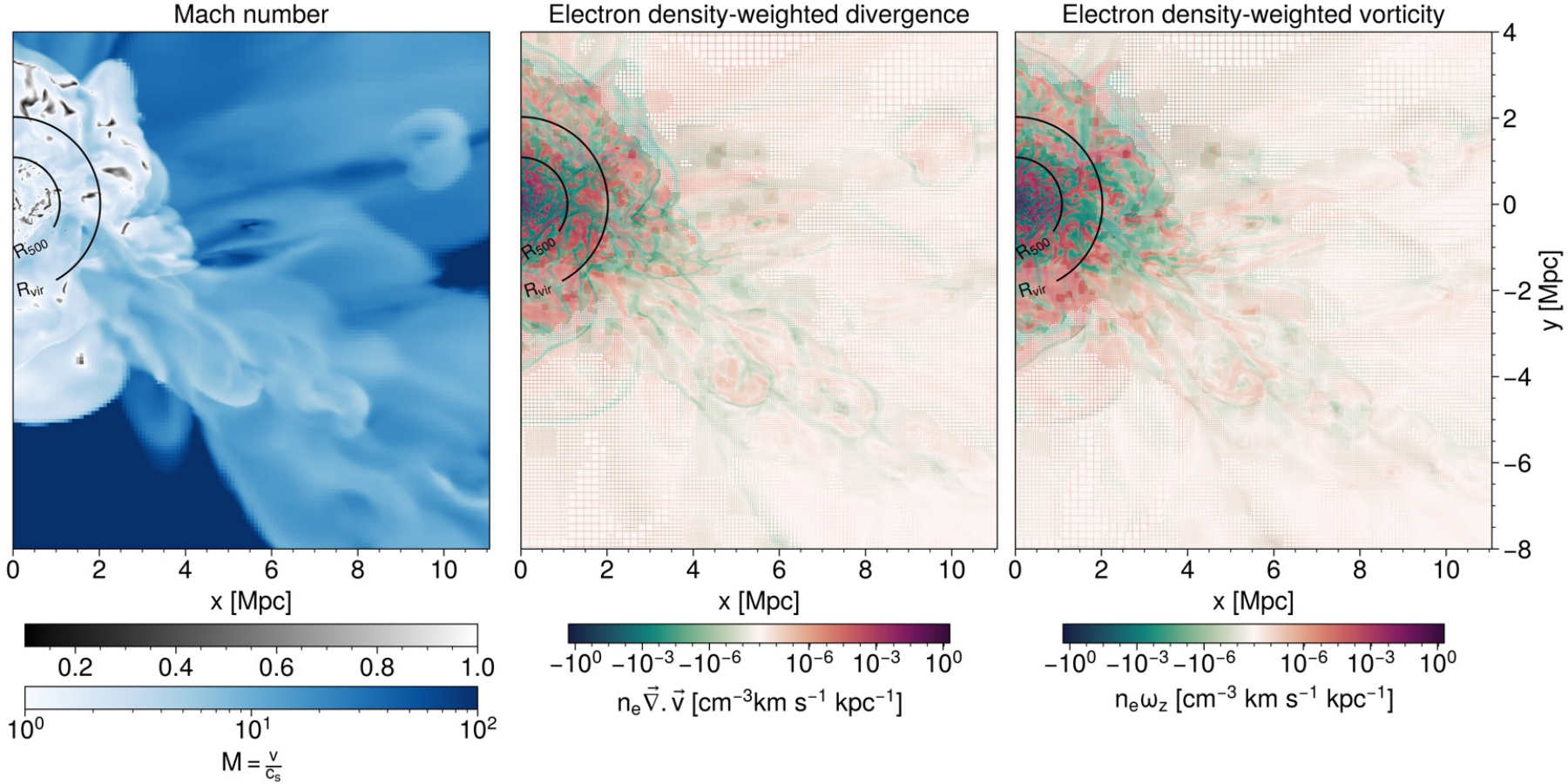}
            
            \caption{Zoom on the background (top) and foreground (bottom) filaments in the longitudinal cut. The maps are $11.061\times12~Mpc$ large with the same resolution as those presented in Fig. \ref{long_cut}. The left panel shows the Mach number, the central panel shows the electron density-weighted divergence, and the right panel shows the z component of the electron density-weighted vorticity. Virgo's $R_{500}$ and virial radius, $R_{vir}$, are shown as black circle arcs on each map.}
            
            \label{app:zoom left mach}
       \end{minipage}
\end{figure*}

\section{Power spectra and their ratios computed from 2~Mpc square maps}
\label{appendix 2Mpc square}

\begin{table*}[h!]
    \centering
    \caption{Best fit of the total power spectrum slope}
    \renewcommand{\arraystretch}{1.3}
    \begin{adjustbox}{width=1\textwidth}
    \begin{tabular}{ c c c c c c c c c }
    \hline \hline
    \multicolumn{3}{c}{Kolmogorov : $P(k)\propto k^{-11/3}$ } & \multicolumn{3}{c}{Burgers : $P(k)\propto k^{-4}$ } & \multicolumn{3}{c}{Fit : $P(k)\propto k^{\alpha}$ } \\
    \hline 
    \multicolumn{9}{c}{Best fit of the total power spectrum slope}\\
    \hline
    \hline
    $\Delta x_{cen}$ (Mpc) & -8 & -6 & -4 & -2 & 0 & 2 & 4 & 6 \\
    \hline 
    $\alpha$ (2~Mpc) & -2.64$\pm$0.04 & -2.97$\pm$0.06 & -2.94$\pm$0.07 & -3.11$\pm$0.06 & -2.96$\pm$0.03 & -3.26$\pm$0.06 & -3.02$\pm$0.07 & -2.70$\pm$0.08  \\
    \hline

    \end{tabular}
    \end{adjustbox}
    \label{app:tab:fit}
\end{table*}

\begin{figure}[h!]
    \centering
    \includegraphics[trim= 300 20 300 20,clip ,width=0.7\linewidth]{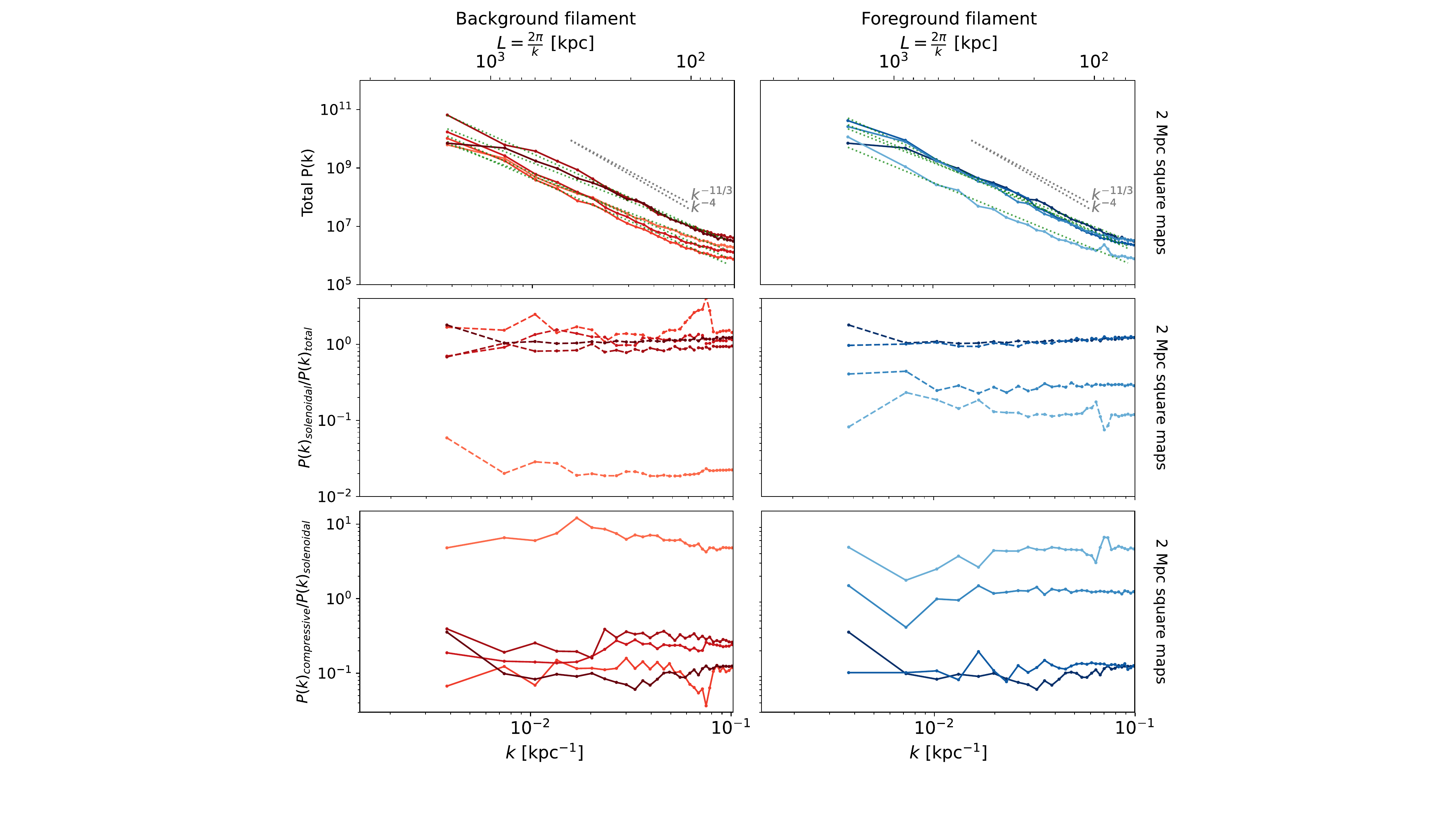}
    \caption{Total power spectra (top) and the ratios of the solenoidal component over the total (middle) and solenoidal over compressive component power spectra (bottom) computed from 2~Mpc square maps, similarly to Figs. \ref{ek_tot_compare}, \ref{ek_ratio_st} and \ref{ek_ratio_cs}.}
    \label{app:ek comp 2Mpc}
\end{figure}

\newpage

\section{Range of validity of the power spectra}

\begin{figure}[h!]
    \begin{minipage}[c]{0.5\textwidth}
    \centering
    \includegraphics[trim=0 0 0 50, clip,width=1\linewidth]{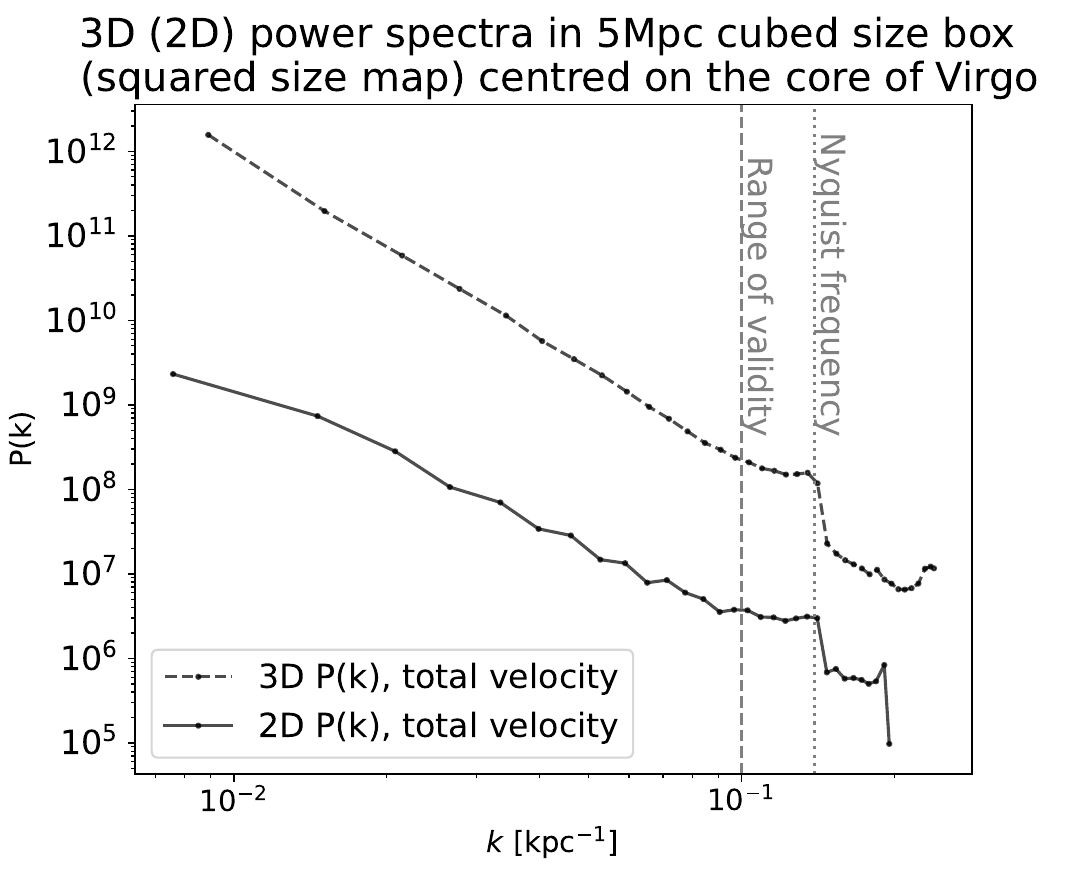}
    \end{minipage}
    \begin{minipage}[c]{0.4\textwidth}
    \caption{Full scale range of the 3D and 2D power spectra computed in the 5~Mpc cubed size box (squared size map) centred on the core of Virgo, similarly to Fig. \ref{2D vs 3D test}. The chosen range of validity and the Nyquist frequency are shown by vertical dashed lines. }
    \label{fig:app:nyq freq}
    \end{minipage}
\end{figure}

\section{Comparison of the 3D and 2D power spectra slopes}
\label{app:2D vs 3D spectra}

\begin{table*}[h!]
    \centering
    \caption{}
    \renewcommand{\arraystretch}{1.3}
    \begin{tabular}{c c c c c c c c c}
    \hline
        \multicolumn{9}{c}{Difference between the 3D and 2D best-fit slope parameter} \\
    \hline \hline
        $\Delta x_{cen} (Mpc)$ & -8 & -6 & -4 & -2 & 0 & 2 & 4 & 6 \\
    \hline
         $\alpha_{3D}-\alpha_{2D}$ & -0.95 & -1.01 & -1.12 & -0.89 & -0.97 & -0.93 & -0.99 & -0.95 \\
    \hline
    \end{tabular}
    \label{tab:app alphas diff}
\end{table*}

\begin{figure}[h!]
    \centering
    \includegraphics[width=0.9\linewidth]{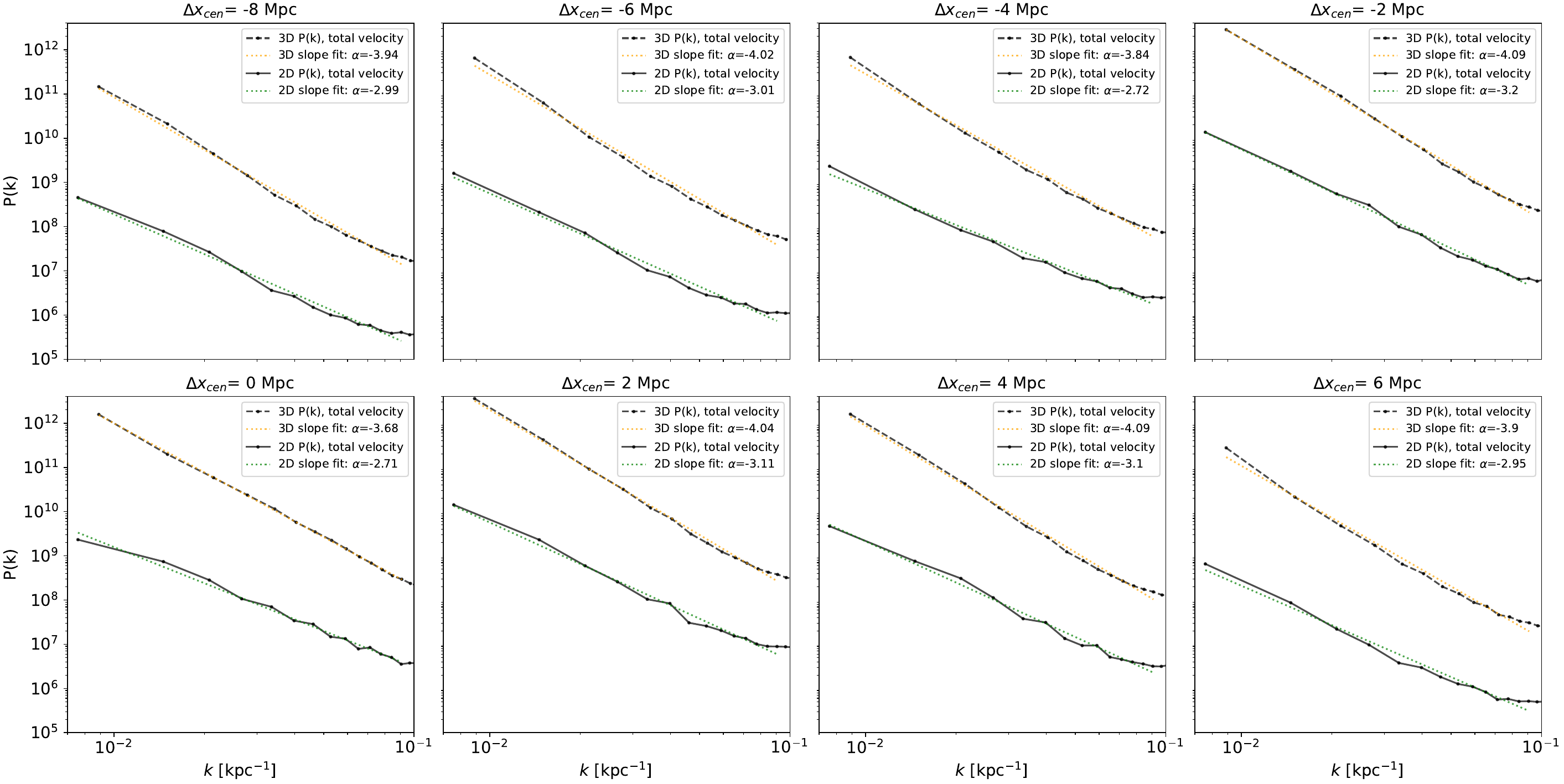}
    \caption{Comparison of the 2D and 3D power spectra slope computed from a 1~Mpc side map or cube with the same centre.}
    \label{2D vs 3D ek comp appendix}
\end{figure}

For each best-fit $\alpha$ parameter, the error, i.e. standard deviation, on the fit is of the order of 1e-3. It is worth noting that the difference between the 3D and 2D slopes is roughly -1, as shown in Tab. \ref{tab:app alphas diff} for all the slices-cube pairs. 

\section{All 2D power spectra}
\label{app:2D_ek_all}

\begin{figure*}[h!]
        \begin{minipage}[s]{1\textwidth}
            \centering
            \includegraphics[width=1\textwidth]{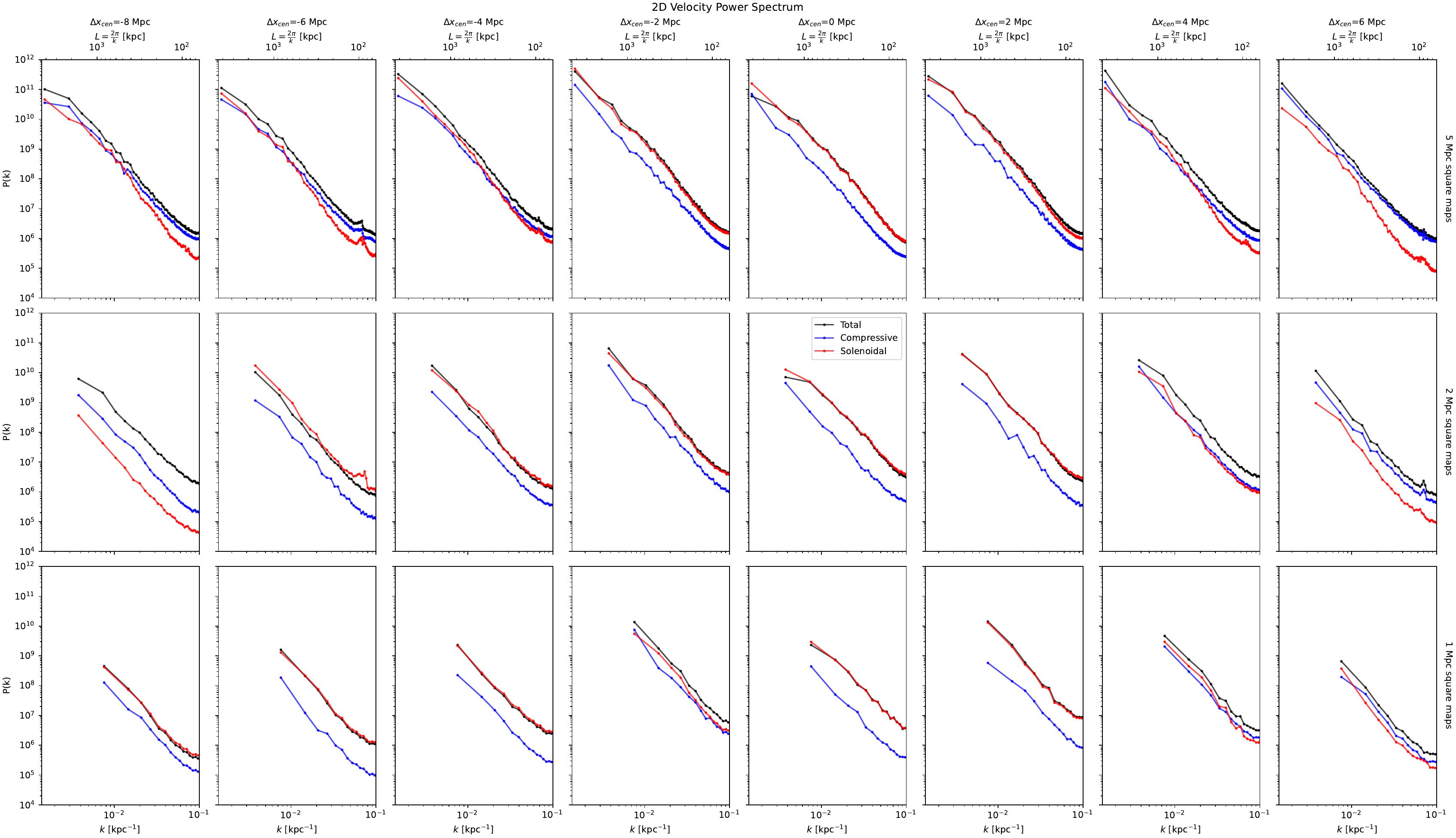}
            \caption{2D power spectrum of the total velocity field (black) and its compressive (blue) and solenoidal (red) components computed from 1 (bottom),2 (middle) and 5~Mpc (top) square maps extracted from the slices. The squares are displayed in the top row of Fig. \ref{decomp_v}, and their centre's position can be found in Tab. \ref{tab:pos}. From left to right, each column presents the power spectra extracted from a given slice at a distance of between -8 and 6~Mpc from Virgo's centre; they are each 2~Mpc apart.}
            \label{ek_2D_all}
       \end{minipage}
\end{figure*}

\end{document}